\documentclass[a4paper,11pt]{article}
\usepackage{amssymb}

\usepackage{graphicx}
\usepackage{epsfig}
\usepackage{amsmath}

\input{tcilatex}

\begin{document}

\baselineskip16pt

\begin{titlepage}

\begin{flushright}
CFP--2005--05\\
ROM2F/2005/10\\
\end{flushright}

\vspace{1cm}

\begin{center}
{\huge \textbf{Unitarity in the Presence of}}\\[0pt]
\vspace{3mm} {\huge \textbf{Closed Timelike Curves}}\\[0pt]

\end{center}

\vspace{8 mm}

\begin{center}
Lorenzo Cornalba$^\dagger$\,\,\,\&\,\, Miguel S. Costa$^\ddagger$

\bigskip
\end{center}

\hspace{0cm}$^{\dagger\,}${\small Dipartimento di Fisica \& INFN,
Universit\'{a} di Roma ``Tor Vergata''}

\hspace{.21cm}{\small Via della Ricerca Scientifica 1, 00133,
Roma, Italy}

\hspace{.21cm}{\small cornalba@roma2.infn.it}

\vspace{5mm}
\hspace{0cm}$^{\ddagger\,}${\small Departamento de F\'{i}sica e
Centro de F\'{i}sica do Porto}

\hspace{0.21cm}{\small Faculdade de Ci\^{e}ncias da Universidade
do Porto}

\hspace{0.21cm}{\small Rua do Campo Alegre, 687, 4169--007 Porto,
Portugal}

\hspace{0.21cm}{\small miguelc@fc.up.pt}
\vspace{5mm}

\begin{abstract}
We conjecture that, in certain cases, quantum dynamics is
consistent in the presence of closed timelike curves. We consider
time dependent orbifolds of three dimensional Minkowski space
describing, in the limit of large AdS radius, BTZ black holes
inside the horizon. Although perturbative unitarity fails, we show
that, for discrete values of the gravitational coupling, particle
propagation is consistent with unitarity. This quantization
corresponds to the quantization of the black hole angular
momentum, as expected from the dual CFT description. Note,
however, that we recover this result by analyzing the physics
inside the horizon and near the singularity. The spacetime under
consideration has no AdS boundary, and we are therefore not using
any assumption regarding a possible dual formulation. We perform
the computation at very low energies, where string effects are
irrelevant and interactions are dominated by graviton exchange in
the eikonal regime. We probe the non--causal structure of
space--time to leading order, but work to all orders in the
gravitational coupling.
\end{abstract}

\end{titlepage}
\newpage \begin{tableofcontents}
\end{tableofcontents}

\section{Introduction}

One of the outstanding difficulties of the
$\mathrm{AdS}/\mathrm{CFT}$ correspondence \cite{Maldacena:1997re}
is to understand physics in the bulk of the $\mathrm{AdS}$ space
in terms of $\mathrm{CFT}$ data. In particular, understanding the
space--time causal structure of black holes is still a fundamental
problem from the view point of the duality. Although one believes,
based on basic properties of the dual $\mathrm{CFT}$, that the
bulk dynamics is well defined, it is fair to say that the quantum
nature of horizons and singularities remains rather mysterious
\cite{Hawking:1976ra}.

The $\mathrm{AdS}_{3}/\mathrm{CFT}_{2}$ case is one of the best studied
examples of the duality, with black hole geometries given by the BTZ metric
\cite{Banados:1992wn,Banados:1992gq}
\begin{equation}
ds^{2}=-N^{2}dt^{2}+N^{-2}dr^{2}+r^{2}\left( d\phi -N_{\phi }dt\right)
^{2}\,,  \label{BTZ}
\end{equation}
where
\begin{equation*}
N^{2}=\frac{1}{\ell ^{2}r^{2}}\left( r^{2}-r_{+}^{\,2}\right) \left(
r^{2}-r_{-}^{\,2}\right) \,,\,\ \ \ \ \ \ \ \ \ \ \ \ \ \ \ \ \ N_{\phi }=
\frac{1}{\ell }\frac{r_{+}r_{-}}{r^{2}}\,.
\end{equation*}
The $\mathrm{AdS}_{3}$ radius is given by $\ell $, and
$r_{+},r_{-}$ are the positions of the outer and inner horizons
determining the mass and the angular momentum of the black hole
\begin{equation*}
M_{\mathrm{bh}}=\frac{\pi M}{4}\left( \frac{r_{+}^{\,2}+r_{-}^{\,2}}{\ell ^{2}}
+1\right) \,,\,\ \ \ \ \ \ \ \ \ \ \ \ \ \ \ \ \ \ J=\frac{\pi M}{2}
\frac{r_{+}r_{-}}{\ell }\,,
\end{equation*}
in terms of the three--dimensional Planck mass\footnote{For notational convenience,
we normalize the Planck mass in terms of the
Newton constant  $G$ as $M^{-1}=2\pi G$.} $M$.

In the dual $\mathrm{CFT}_{2}$ description, these black holes correspond to
states with \cite{Strominger:1997eq}
\begin{equation*}
L_{0}+\tilde{L}_{0}=\ell M_{\mathrm{bh}},\,\ \ \ \ \ \ \ \ \ \ \ \ \ \ \
L_{0}-\tilde{L}_{0}=J,
\end{equation*}
where $L_{0},\tilde{L}_{0}$ are the Virasoro zero modes. Moreover, since
the $\phi $ circle is a non--contractible loop in spacetime,
in the presence of fermions one needs to choose a spin structure.
Usually one considers periodic boundary conditions for the
fermions, which allow for a covariantly constant spinor in the
extremal case $r_{+}=r_{-}$ and which give a state in the Ramond
sector of the $\mathrm{CFT}_{2}$. One may also choose antiperiodic
boundary conditions, which describe a non--supersymmetric state in
the NS sector of the $\mathrm{CFT}_{2}$. From this point of view,
the spin eigenvalue $J$ is naturally quantized in half integral
units
\begin{equation}
2J\in \mathbb{Z}\ .  \label{Jquantization}
\end{equation}
On the other hand, from a purely gravitational view point, the
quantization of the angular momentum is rather mysterious.
Classically, $J$ is a continuous parameter, and the usual
arguments leading to (\ref {Jquantization}) rely on the asymptotic
symmetries of quantum gravity on $\mathrm{AdS}_{3}$ \cite{Brown:1986nw}, and therefore
implicitly on the existence of a dual $\mathrm{CFT}_{2}$.

An intriguing property of the BTZ black holes is the existence of closed
causal curves (CCC's) in the geometry. In fact, these holes are quotients of
$\mathrm{AdS}_{3}$ by the action of a specific isometry parameterized by $%
r_{+},r_{-}$, and the identification creates CCC's located in the
region inside the inner horizon. The black hole has a
chronological singularity where the
generator of the orbifold isometry becomes null. Therefore, if we
ignore the dual $\mathrm{CFT}$ description, we naively expect that
quantum gravity in the BTZ geometry presents pathologies due to
the existence of these CCC's. In particular, one would a priori
expect violations of unitarity, which would undermine the possible
existence of an $S$ matrix.

\begin{figure}
\centering\psfig{figure=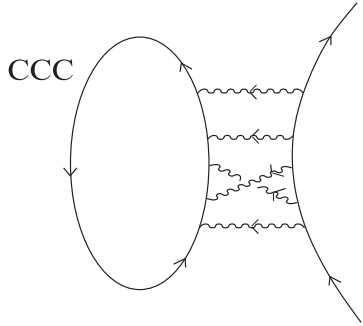,width=5cm}
\caption{\small{Leading correction to the free propagation of a
scalar field due to gravitational interactions with virtual
particles winding closed causal curves.}}
\label{FigEikonal}
\end{figure}

By studying quantum field theory in the flat space limit $\ell \rightarrow
\infty $ of the BTZ geometry, we shall show that the quantization condition (%
\ref{Jquantization}) can be obtained by demanding that quantum
propagation of fields is consistent with unitarity, even in the
presence of CCC's. The argument will use very limited information
about the underlying quantum gravity. In particular, we shall not
use any string theoretic arguments, since we shall work at
energies well below the string scale, where $\alpha ^{\prime }$
corrections should be negligible. More specifically, we will
consider corrections to free propagation of scalar fields due to
interactions with particles winding around the closed timelike
direction, as shown in Figure \ref{FigEikonal}. By carefully
choosing the quantum numbers of the external states, we will show,
using general arguments \cite{'tHooft:1987rb}, that the interaction is
dominated by graviton exchange in the eikonal regime. In this
kinematical regime, one has enough control over the resummation of
the perturbative series determining the gravitational interaction,
and one can recover the quantization condition
(\ref{Jquantization}) by enforcing unitarity.

For a given BTZ black hole with geometric parameters
$r_{+},r_{-}$, one could in principle start by considering quantum
gravity at arbitrary values of the gravitational coupling
$M^{-1}$. However, the presence of CCC's breaks unitarity order by
order in the coupling constant \cite{Boulware:1992pm}, suggesting that one is not free to
choose $M^{-1}$ at will and that one needs to resum the
perturbation series. The quantization condition
(\ref{Jquantization}), which follows from unitarity, can then be
interpreted as fixing the gravitational coupling $M^{-1}$ at some
specific values.

The usual belief regarding the dynamics inside the horizons of a BTZ black hole is that the
geometry (\ref{BTZ}) will be modified due to the classical instability of
the inner horizon \cite{Simpson}
and also due to divergences of the quantum stress--energy tensor at polarization surfaces
in the region inside the inner horizon \cite{Kim:1991mc}.
We propose, on the other hand, that the geometry
remains unaltered and non--causal, but the propagation of states is
strongly modified by quantum effects in the inner region and is consistent
whenever (\ref{Jquantization}) is satisfied. It is tempting to
speculate that such modification of the
dynamics inside the horizons is responsible for the reduction of
degrees of freedom associated to the black hole entropy, which is not
proportional to the volume of the black hole but to its horizon area.

\section{The orbifold}

\begin{figure}
\centering\psfig{figure=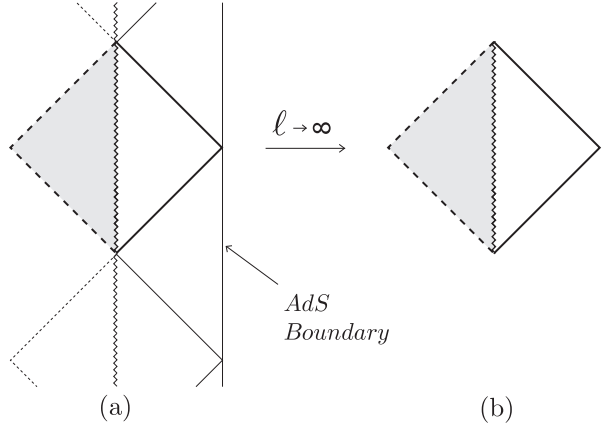,width=8cm}
\caption{\small{Penrose diagram of the extremal BTZ black hole
(a). The shaded area represents the region behind the
chronological singularity, where closed causal curves are present.
In the limit $\ell\to\infty$, $J$ fixed, one focuses on the region
inside the black hole horizon and obtains a flat space orbifold
with Penrose diagram (b).}} \label{FigExtremal}
\end{figure}

For simplicity we focus on the extremal black hole, with $r_{+}=r_{-}$ and
Penrose diagram given in Figure \ref{FigExtremal}a, which is a quotient of
$\mathrm{AdS}_{3}$ space. In the flat space $\ell \rightarrow \infty $ limit,
keeping the energy scale
\begin{equation*}
E=\frac{\ell }{\left( 2\pi r_{+}\right) ^{2}}
\end{equation*}
fixed, the region inside the black hole horizon becomes an
orbifold of flat Minkowski space $\mathbb{M}^{3}/e^{\kappa }$,
introduced in \cite{Cornalba:2003kd}. Choosing coordinates
$x^{\pm},x$ on $\mathbb{M}^{3}$, such that the metric is
\begin{equation*}
ds^{2}=-2dx^{+}dx^{-}+dx^{2}\,,
\end{equation*}
the orbifold generator $\kappa $ is the Killing vector
\begin{equation*}
\kappa =i\left( L_{+x}+E^{-1}K_{-}\right) =-\left( x^{-}\partial
_{x}+x\partial _{+}\right) +E^{-1}\partial _{-}\,,
\end{equation*}
where $L_{ab}$, $K_{a}$ are, respectively, the generators of Lorentz transformations and
translations, and where $E$ parameterizes inequivalent orbifolds.
For a detailed derivation of the $\ell \rightarrow \infty $
limiting procedure, we refer the reader to appendix A.

Under the change of coordinates
\begin{eqnarray}
x^{+} &=&y^{+}-Eyy^{-}+\frac{E^{2}}{6}\left( y^{-}\right) ^{3}\,\,,  \notag
\\
x^{-} &=&y^{-}\,\,,  \label{coordchange} \\
x &=&y-\frac{E}{2}\left( y^{-}\right) ^{2}\,,  \notag
\end{eqnarray}
the metric becomes
\begin{equation}
ds^{2}=-2dy^{+}dy^{-}+2Ey\left( dy^{-}\right) ^{2}+dy^{2}  \label{eq100}
\end{equation}
and the Killing vector
\begin{equation*}
\kappa =\frac{1}{E}\frac{\partial }{\partial y^{-}}\,.
\end{equation*}
The direction $y^{-}$ is therefore compact with
period
\begin{equation*}
y^{-} \sim y^{-}+\frac{1}{E}\,.
\end{equation*}
Note that the form of the metric (\ref{eq100}) can be directly
obtained from the BTZ metric (\ref{BTZ}) for $r_{+}=r_{-}$, by
setting $y=2\pi ^{2}Er^{2}$, $2\pi y^{+}=t$, $2\pi E y^- = \phi$,
and then by taking the limit $\ell \rightarrow \infty $ described
above. Re--expressing the metric (\ref{eq100}) as
\begin{equation*}
ds^{2}=-\frac{\left( dy^{+}\right) ^{2}}{2Ey}+dy^{2}+2Ey\left( dy^{-}-
\frac{dy^{+}}{2Ey}\right) ^{2}\,,
\end{equation*}
we can easily draw the corresponding Penrose diagram as in Figure
\ref{FigExtremal}b, showing that we are focusing on the region inside the horizon of
the extremal BTZ black hole. In the non--extremal case a
similar large $\ell $ limit leads to the shift--boost orbifold of $\mathbb{M}^{3}$,
which focuses on the region inside the outer horizon \cite{Cornalba:2002fi}.

The quantization of the BTZ black hole angular momentum (\ref{Jquantization})
becomes, in the flat space limit, the condition
\begin{equation}
\ 2J=\frac{M}{4\pi E}\in \mathbb{Z\,}.  \label{Magic}
\end{equation}
In this case, on the other hand, one cannot justify this quantization
condition with arguments relying on asymptotic symmetries and on the
existence of a dual CFT. In fact, the Minkowski space orbifold just
described focuses on the region inside the horizons, and the asymptotic
$\mathrm{AdS}$ boundary is no longer part of the geometry.

An independent way of deriving (\ref{Magic})
is to embed the orbifold in string theory, by considering Type II
strings on $\mathbb{M}^{3}/e^{\kappa }\times \mathbb{T}^{7}$ \cite{Cornalba:2003kd}. After a
sequence of dualities, the geometry becomes that of an orientifold $O8$
plane \cite{Cornalba:2002nv}. From this point of view, (\ref{Magic}) results from the fact that
$8$--dimensional RR charged objects have charges quantized in units of the
$D8$--brane charge. Although these arguments rely on the low--energy
supergravity description of the system, they lead to the same condition.

It is the main point of this paper to derive the quantization
condition (\ref{Magic}) purely within the framework of quantum
field theory in the presence of gravitational interactions. From
this perspective, (\ref{Magic}) is obtained by requiring unitarity
in the space $\mathbb{M}^{3}/e^{\kappa }$, which possess CCC's.
Hence we see that unitarity in the presence of CCC's is related to
charge quantization in dual descriptions of the system. The
mechanics that protects chronology is rather different than that
proposed by Hawking \cite{Hawking:1991nk}, which is based on a
large backreaction due to UV effects.

For other studies of orbifolds with CCC's, where an analogous study of unitarity can
in principle be done, see [15--32].

\subsection{Geometry}

\begin{figure}
\centering\psfig{figure=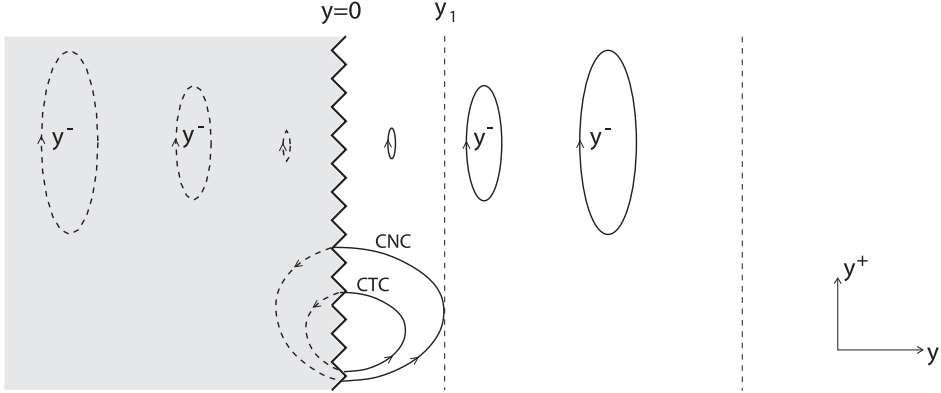,width=12.5cm} \caption{\small{A
representation of the basic features of the orbifold geometry. The
null Killing direction $y^+$ and the transverse direction $y$ are drawn
on the plane. The compact $y^-$ circle is spacelike for positive
$y$ and timelike in the shaded region after the chronological
singularity. Represented are also the first polarization surfaces
at $y_w=w^2/(24E)$. Finally we show a closed timelike curve and a closed null curve.
The latter starts and ends on the first polarization surface.}} \label{FigCCC}
\end{figure}

We now analyze, in more detail, the geometry of the orbifold $\mathbb{M}^{3}/e^{\kappa }$.
First let us note that the square--norm of $\kappa $ is
given by
\begin{equation*}
\kappa ^{2}=\frac{2y}{E}\,.
\end{equation*}
The compact $y^{-}$ circle is space--like for $y>0$ and timelike
for $y<0$. Therefore, the geometry has CCC's. It is simple to show
that all CCC's must go in the region with $y<0$. To prove this
fact, assume that we have a CCC parameterized by $y^{a}\left(
\lambda \right) $ with $\lambda \in \left[ 0,1\right] $. Since
$y^{+}\left( 0\right) =y^{+}\left( 1\right) $, the function
$y^{+}\left( \lambda \right) $ must achieve an extremum for some
$\bar{\lambda}\in \left( 0,1\right) $. At $\lambda
=\bar{\lambda}$, the metric (\ref{eq100}) reduces to $2Ey\left(
dy^{-}\right) ^{2}+dy^{2}$, which is positive--definite for $y>0$.
Therefore, the curve can be timelike or null only if $y\left(
\bar{\lambda}\right) \leq 0$. A schematic representation of these
basic features of the geometry is shown in Figure \ref{FigCCC}.

A particularly interesting class of closed curves are the ones obtained from
geodesics in the covering space $\mathbb{M}^{3}$ connecting points which are
related by the action of the orbifold group. Denoting, for notational
convenience, the orbifold group generator by
\begin{equation*}
\Omega =e^{\kappa },
\end{equation*}
it is a simple matter to show, starting from (\ref{coordchange}),
that the continuous transformation $y^{-}\rightarrow y^{-}+s/E$
generated by $\Omega ^{s}$ reads, in the coordinates $x^{a}$,
\begin{eqnarray}
\left( \Omega ^{s}\mathbf{x}\right)^{+}& =& x^{+}-sx+\frac{s^2}{2}\,x^{-}+
\frac{s^3}{6E}\,,  \notag \\
\left( \Omega ^{s}\mathbf{x}\right)^{-}& =& x^{-}+\frac{s}{E}\,,
\label{transf} \\
\left( \Omega ^{s}\mathbf{x}\right)&  =& x-sx^{-}-\frac{s^2}{2E}\,.
\notag
\end{eqnarray}
Therefore, the geodesic distance squared
$\left( \Omega ^{w}\mathbf{x}-\mathbf{x}\right) ^{2}$ between a point $\mathbf{x}$ and its
$w$--th image ($w\in \mathbb{Z}$) is given by
\begin{equation*}
\frac{2w^{2}}{E}\left( y-\frac{w^{2}}{24E}\right) \,,\,
\end{equation*}
and becomes null at the so--called polarization surfaces
\begin{equation*}
y=\frac{w^{2}}{24E}\,.
\end{equation*}
For $y<w^{2}/(24E)$ the geodesic from $\mathbf{x}$ to $\Omega ^{w}\mathbf{x}$
is a closed timelike curve. The polarization surfaces are also represented in
Figure \ref{FigCCC}.

\subsection{Particle wavefunctions\label{wavefunctions}}

In order to analyze the propagation of scalar fields in the orbifold
geometry, we must first find a convenient basis of functions invariant under
the orbifold action. It is useful to consider first the transformation of
the plane waves
\begin{equation*}
\phi _{\mathbf{k}}(\mathbf{x}) =e^{i\mathbf{k}\cdot \mathbf{x}}
\end{equation*}
under the action of $\Omega $. Using (\ref{transf}), it is simple to show
that
\begin{equation}
\phi _{\mathbf{k}}\left( \Omega ^{-s}\mathbf{x}\right) =\phi _{\Omega ^{s}
\mathbf{k}}(\mathbf{x})\, e^{-i\varphi (\mathbf{k},s)
},  \label{pwave}
\end{equation}
where the transformed momentum $\Omega ^{s}\mathbf{k}$ is given by
\begin{eqnarray}
\left( \Omega ^{s}\mathbf{k}\right) _{+} &=&k_{+}\,,
\notag \\
\left( \Omega ^{s}\mathbf{k}\right) _{-} &=&k_{-}+sk+\frac{s^{2}}{2}\,k_{+}\,,
\notag \\
\left( \Omega ^{s}\mathbf{k}\right) &=&k+sk_{+}\,.
\label{ptransf}
\end{eqnarray}
The above phase $\varphi (\mathbf{k},s)$ is given by
\begin{equation*}
\varphi (\mathbf{k},s) =\frac{1}{E}\left( \frac{s^{3}}{6}\,k_{+}
+sk_{-}+\frac{s^{2}}{2}\,k\right) ,
\end{equation*}
and satisfies $\varphi (\mathbf{k},s+t) =\varphi (\mathbf{k},s)
+\varphi (\Omega^{s}\mathbf{k},t)$.

We can construct, starting from any plane wave $\phi _{\mathbf{k}}(\mathbf{x})$,
a function on $\mathbb{M}^{3}$ which is invariant under
$\Omega $ and which is an eigenvector of $\kappa $ with eigenvalue $2\pi in$,
where $n\in \mathbb{Z}$. This is given by the integral representation
\begin{equation}
\int ds\,e^{2\pi ins}\phi _{\mathbf{k}}\left( \Omega ^{-s}\mathbf{x}\right)
=\int ds\;\phi_{\Omega^{s}\mathbf{k}}(\mathbf{x}) \,e^{2\pi ins-i\varphi(\mathbf{k},s) }\,.
\label{intrep}
\end{equation}
Since $\left( \Omega ^{s}\mathbf{k}\right) ^{2}=\mathbf{k}^{2}$,
and since $\left( \Omega ^{s}\mathbf{k}\right) _{+}=k_{+}$, the
above functions are automatically eigenvectors of the Laplacian
$\square $ and of the momentum operator $K_{+}$. We
then choose, as a convenient basis for the invariant functions on $\mathbb{M}%
^{3}/e^{\kappa }$, eigenfunctions $V_{\lambda ,p_{+},p_{-}}(\mathbf{x})$
of the commuting operators $\square $, $K_{+}$ and $\kappa $, with
eigenvalues
\begin{eqnarray}
\square &=&\lambda \,,  \notag \\
K_{+} &=&p_{+}\,,  \label{diag} \\
-iE\kappa &=&p_{-}\,,  \notag
\end{eqnarray}
where $p_{-}$ is related to the eigenvalue $n$
by\footnote{Throughout the paper, we use $\mathbf{k},\mathbf{q}$
to denote covering space momenta -- \textit{i.e} eigenvalues of
$-i\partial_\mathbf{x}$. We reserve $p_\pm$ for the momenta in the
$\mathbf{y}$ coordinates $-i\partial_{y^\pm}$}
\begin{equation*}
p_{-}=2\pi E\,n\,.
\end{equation*}
Choosing, in (\ref{intrep}), the momentum $\mathbf{k}=\left( p_{+},\lambda
/2p_{+},0\right) $, we immediately obtain the explicit representation
\begin{eqnarray}
V_{\lambda ,p_{+},p_{-}}(\mathbf{x}) &=&\frac{1}{\left|
p_{+}\right| }\int dk\,e^{\,i\left( p_{+}x^{+}+k_{-}x^{-}+kx\right) }\,\times
\label{inv1} \\
&&\times \exp \frac{i}{2Ep_{+}^{\,2}}\left[ \left( 2p_{+}p_{-}-\lambda \right)
\,k-\frac{k^{3}}{3}\right] ,  \notag
\end{eqnarray}
where we have changed the integration variable to $k=sp_{+}$, and where $%
k_{-}$ is given by the on--shell condition
\begin{equation*}
k_{-}=\frac{k^{2}+\lambda }{2p_{+}}\,.
\end{equation*}

The sector with $p_{+}=0$ clearly requires a separate treatment. In this
case, (\ref{ptransf}) implies that the component $k=\left( \Omega ^{s}\mathbf{k}\right) $
is invariant under $\Omega $, and we can choose eigenfunctions $%
V_{p,p_{-}}(\mathbf{x})$ with
\begin{eqnarray*}
K &=&p\,, \\
-iE\kappa &=&p_{-}\,,
\end{eqnarray*}
where $K$ is the momentum operator $-i\partial_x$. Choosing, in
(\ref{intrep}), the momentum $\mathbf{k}=\left( 0,0,p\right) $,
and as integration variable $k_{-}=sp$, we obtain the integral
representation
\begin{equation}
V_{p,p_{-}}(\mathbf{x}) =\frac{1}{\left| p\right| }\int
dk_{-}\,e^{\,i\left( k_{-}x^{-}+px\right) }\,\exp \frac{i}{pE}\left[
p_{-}k_{-}-\frac{\left( k_{-}\right) ^{2}}{2}\right] ,  \label{inv2}
\end{equation}
which has $p_{+}=0$ and $\lambda =-p^{2}$.

The functions $V_{\lambda ,p_{+},p_{-}}$ and $V_{p,p_{-}}$ represent a
complete basis of invariant functions on $\mathbb{M}^{3}/e^{\kappa }$, and
transform, under conjugation, as
\begin{eqnarray*}
V_{\lambda ,p_{+},p_{-}}^{\,\star }(\mathbf{x}) &=&V_{\lambda,-p_{+},-p_{-}}(\mathbf{x}) \,, \\
V_{p,p_{-}}^{\,\star}(\mathbf{x}) &=&V_{-p,-p_{-}}(\mathbf{x}) \,.
\end{eqnarray*}

It is useful to consider the above functions in the coordinates $y^{\pm },y$.
The operators in (\ref{diag}) are given, in these coordinates, by
\begin{eqnarray*}
\square &=&-2\partial _{y^{+}}\partial _{y^{-}}-
2Ey\partial_{y^{+}}^{\,2}+\partial _{y}^{\,2}\,, \\
K_{+} &=&-i\partial _{y^{+}}\,, \\
-iE\kappa &=&-i\partial _{y^{-}}\,.
\end{eqnarray*}
We can then use separation of variables to find functions
satisfying (\ref{diag}) of the form $f(y)
\,e^{\,i(p_{+}y^{+}+p_{-}y^{-}) }$, where $f(y)$ solves
\begin{equation}
\left( 2p_{+}p_{-}+2Ep_{+}^{\,2}y+\frac{d^{2}\,}{dy^{2}}-\lambda \right)\,f(y) =0\,.
\label{ODE}
\end{equation}
Defining the dimensionless variable
\begin{equation}
z=\left( 2Ep_{+}^{\,2}\right) ^{\frac{1}{3}}\left( y-y_{0}\right) \,,\ \ \
\ \ \ \ \ \ \ \ \ \ \ \ \ \ \ \ \ \ \ \ \
\left( y_{0}=\frac{\lambda - 2p_{+}p_{-}}{2Ep_{+}^{\,2}}\right)  \notag
\end{equation}
the above differential equation simplifies to
$d^{2}f/dz^{2}+zf=0$, which describes in quantum mechanics a zero
energy particle subject to a linear potential. The solutions are
the Airy functions $\mathrm{Ai}(-z)$ and $\mathrm{Bi}(-z)$, which
are, respectively, exponentially damped and exponentially growing
in the $z<0$ region. This region corresponds mostly to negative
$y$, where the Killing vector $\kappa $ is time--like. In the
sequel, we shall consider the normalizable solution
$\mathrm{Ai}\left( -z\right) $ which corresponds to the integral
representation (\ref{inv1}). Using the representation of the Airy
function
\begin{equation}
\mathrm{Ai}(-z)=\frac{1}{2\pi }\int dt\,\,e^{\,i\left( zt-\frac{t^{3}}{3}\right)}\,,  \notag
\end{equation}
it is a matter of computation to show, using (\ref{coordchange}) and (\ref
{inv1}), that
\begin{equation*}
V_{\lambda ,p_{+},p_{-}}(\mathbf{y}) = 2\pi \left(\frac{2E}{\left|
p_{+}\right| }\right) ^{\frac{1}{3}}\mathrm{Ai}(-z) \,e^{\,i(
p_{+}y^{+}+p_{-}y^{-}) }\,.
\end{equation*}
The case $p_{+}=0$ is simpler since in the differential equation
(\ref{ODE}) the linear potential term is absent. The solutions are
simply plane waves $f(y) \propto e^{\pm ipy}$, with
$p^{2}=-\lambda $. In fact, we can explicitly integrate (\ref
{inv2}) and obtain
\begin{equation}
V_{p,p_{-}}(\mathbf{y}) =\sqrt{\frac{2\pi
E}{ip}}e^{\,i(py+p_{-}y^{-})}\,. \label{exp2}
\end{equation}

Finally, the inner product of the wave functions is given by
\begin{eqnarray}
\int_{\mathcal{F}}d^{3}\mathbf{x}\,\,V_{\lambda ,p_{+},p_{-}}(\mathbf{x})\,
V_{\lambda ^{\prime },p_{+}^{\prime },p_{-}^{\prime }}^{\,\star}(\mathbf{x})
&=&16\pi ^{3}\,\delta \left( \lambda -\lambda
^{\prime }\right) \delta \left( p_{+}-p_{+}^{\prime }\right) \delta
_{p_{-}-p_{-}^{\prime }}\,\,,  \notag \\
\int_{\mathcal{F}}d^{3}\mathbf{x}\,\,V_{p,p_{-}}(\mathbf{x})\,
V_{p^{\prime },p_{-}^{\prime }}^{\,\star }(\mathbf{x})
&=&\frac{4\pi ^{2}}{\left| p\right|}\,T\,\delta \left( p-p^{\prime }\right) \delta
_{p_{-}-p_{-}^{\prime }}\,\,\,,  \label{orth}
\end{eqnarray}
where $T=\int dy^{+}=2\pi \delta \left( p_{+}=0\right) $ is the volume in
the $y^{+}$ direction and $\mathcal{F}\subset \mathbb{M}^{3}$ is a
fundamental domain of the orbifold (we can consider $\mathcal{F}$ to be, for
concreteness, the region $0<y^{-}<1/E$). A simple way to prove the above
expressions is to use the integral representations (\ref{inv1}) and (\ref{inv2}) and to
extend the integration region from $\mathcal{F}$ to $\mathbb{M}%
^{3}$. One then uses the orthogonality of the plane waves $\phi _{\mathbf{k}%
}\left( \mathbf{x}\right) $ to derive the desired result. The overcounting
due to the extension of the integration region can be taken into account by
substituting, at the end of the computation, the Dirac $\delta $--function
with the Kronecher symbol for the discrete Ka\l u\.{z}a--Klein charge as
follows
\begin{eqnarray}
\delta \left( n-n^{\prime }\right) &\rightarrow &\delta _{n-n^{\prime }}\,,
\label{overcount} \\
2\pi\, \delta \left( p_{-}-p_{-}^{\prime }\right) &\rightarrow &
E^{-1}\delta_{p_{-}-p_{-}^{\prime }}\,.  \notag
\end{eqnarray}
The correctness of the normalization in the above expression can be checked,
for instance, by computing (\ref{orth}) directly in the $y$--coordinates,
using the explicit expression (\ref{exp2}).

\section{Feynman rules\label{FeynRules}}

We shall start our investigation of quantum field theory in the orbifold
space by considering, as a simple toy model, a scalar field $\Phi $ with a
cubic coupling, described by the action
\begin{equation}
\int_{\mathcal{F}}d^{3}\mathbf{x}\;
\left[ \frac{1}{2}\,\Phi \left(\square-m^{2}\right)\Phi +\frac{g}{6}\,\Phi ^{3}\right] \,.
\label{action}
\end{equation}
We shall analyze the perturbative expansion of this theory in some detail,
since some of the basic features of field and string theory on $\mathbb{M}%
^{3}/e^{\kappa }$ can be already understood in this simple setting. Later on
we will be interested in the free theory of a scalar field coupled to
gravity.

\begin{figure}
\centering\psfig{figure=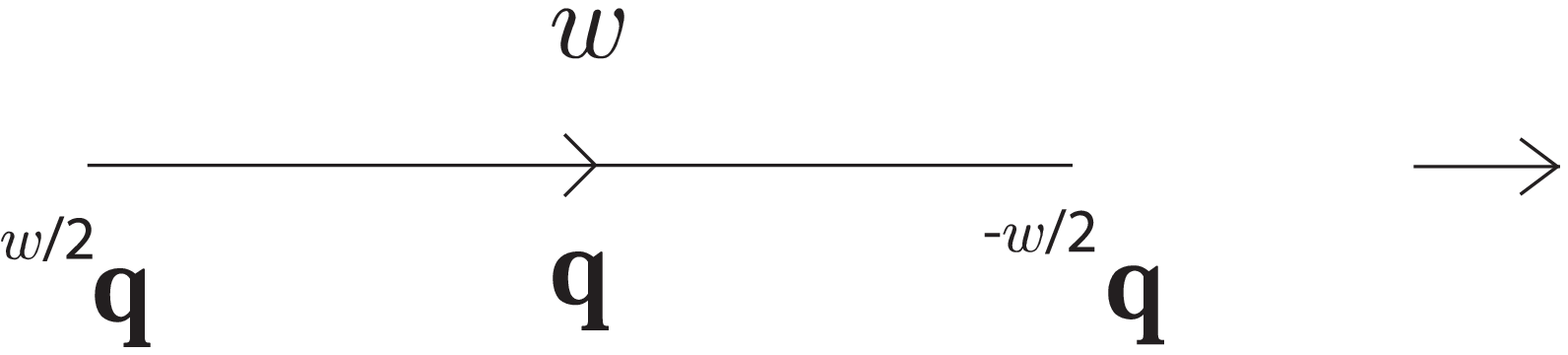,width=10cm}
\begin{picture}(0,0)(0,0)
\put(-297,8){{$\Omega$}} \put(-202,8){{$\Omega$}}
\put(-110,18){\Large{$\frac{-i}{{\bf q}^2 + m^2
-i\epsilon}\;e^{\,\frac{i}{E} \left({w}q_- +
\frac{{w}^3}{24}\,q_+\right)}$}}
\end{picture}
\caption{\small{Scalar propagator for a particle winding $w$ times
the compact $y^-$ direction. The incoming and outgoing momenta are
related by the action of the orbifold generator and the usual
propagator is multiplied by a momentum dependent phase.}}
\label{FigPropagator}
\end{figure}

Consider first the scalar propagator, which is simply given by the method of
images. Denoting the Feynman propagator in the covering space by
\begin{equation*}
\Delta \left( \mathbf{x},\mathbf{x}^{\prime }\right) =
\int \frac{d^{3} \mathbf{q}}{\left( 2\pi \right) ^{3}}\,
\frac{-i}{\mathbf{q}^{2}+m^{2}-i\epsilon }\,
e^{\,i\mathbf{q}\left( \mathbf{x}-\mathbf{x}^{\prime }\right) }\,,
\end{equation*}
we can write the full propagator as a sum
\begin{equation*}
\left\langle\Phi\left(\mathbf{x}\right)\Phi\left(\mathbf{x}^{\prime}\right)\right\rangle
=\sum_{w\in \mathbb{Z}}\Delta\left(\Omega ^{w}\mathbf{x},\mathbf{x}^{\prime }\right) \,.
\end{equation*}
Lorentz invariance implies that $\Delta \left( \Omega ^{s}\mathbf{x},\Omega
^{s}\mathbf{x}^{\prime }\right) =\Delta \left( \mathbf{x},\mathbf{x}^{\prime
}\right) $, and therefore we can write the summand $\Delta \left( \Omega ^{w}\mathbf{x}%
,\mathbf{x}^{\prime }\right) =\Delta \left( \Omega ^{w/2}\mathbf{x},\Omega
^{-w/2}\mathbf{x}^{\prime }\right) $ symmetrically as
\begin{eqnarray}
&&\hspace{-0.5in}\int \frac{d^{3}\mathbf{q}}{\left( 2\pi \right)
^{3}}\, \frac{-i}{\mathbf{q}^{2}+m^{2}-i\epsilon }\;
\phi_{\mathbf{q}}( \Omega ^{w/2}\mathbf{x})
\phi_{\mathbf{q}}^{\,\star}(\Omega^{-w/2}\mathbf{x}^{\prime})
\label{eq1100} \\
&=&\int \frac{d^{3}\mathbf{q}}{\left(2\pi\right)^{3}}\,
\frac{-i}{\mathbf{q}^{2}+m^{2}-i\epsilon }\;
e^{\,\frac{i}{E}\left( wq_{-}+\frac{w^3}{24}\,q_{+}\right)}\,
\phi_{\Omega^{-w/2}\mathbf{q}}\left( \mathbf{x}\right)
\phi_{\Omega^{w/2}\mathbf{q}}^{\,\star}\left(\mathbf{x}^{\prime}\right) \,.
\notag
\end{eqnarray}
In the last equation we have used equation (\ref{pwave}) to obtain the phase
\begin{equation*}
\varphi\left(\mathbf{q},\frac{w}{2}\right)-\varphi\left(\mathbf{q},-\frac{w}{2}\right)\,.
\end{equation*}
>From equation (\ref{eq1100}) we deduce that, in Fourier space, a
scalar propagator is labeled by a momentum $\mathbf{q}$ and a
winding number $w$. The propagator is then given by
\begin{equation*}
\frac{-i}{\mathbf{q}^{2}+m^{2}-i\epsilon }\;
e^{\,\frac{i}{E}\left( wq_{-}+\frac{w^3}{24}\,q_{+}\right) }\,.
\end{equation*}
Moreover, as we move along the propagator, the momentum gets
transformed under the action of the orbifold group element $\Omega
^{-w}$. Therefore, the incoming momentum along the line is $\Omega
^{w/2}\mathbf{q}$ and the outgoing one is $\Omega
^{-w/2}\mathbf{q}$, as shown in Figure \ref{FigPropagator}. We
have made an explicit choice of $i\epsilon$ prescription for the
propagator, which is implicitly a choice of vacuum for the
spacetime under consideration. This choice is canonical in
orbifold theories, and for BTZ black holes corresponds to the
usual Hartle-Hawking vacuum.

We are now ready to state the Feynman rules for computing the
amputated $n$--point amplitude of a given connected graph $G$. We
are not going to give a formal proof of these rules, since it is
simple but notationally quite cumbersome. We hope that the reader
with some familiarity with the standard techniques of field theory
can convince (her)himself of the validity of the rules below.

\begin{enumerate}
\item[{i)}]  First assign winding numbers $w_{i}$ to all internal propagators of $%
G $. This defines a $1$--cocycle for the graph, and therefore an element of
the cohomology group $\omega \in H^{1}\left( G\right) $. We can think of
propagators in the graph literally as particle propagation. The numbers $w_{i}$
define how many times the particle winds around the compact $%
y^{-}$ circle as one goes around a loop of $G$. For the element $\omega \in
H^{1}\left( G\right) $, the intermediate results of the computation will
depend on the specific choice of representative $w_{i}\left( \omega \right) $%
, but the final result will only depend on the class $\omega $. We should
consider as distinct only choices of windings $w_{i}$ corresponding to
different classes in $H^{1}\left( G\right) $. As an illustration, Figure
\ref{FigLoopGraph} shows the choices of winding numbers and classes for a given
two--loop graph.

\begin{figure}
\centering\psfig{figure=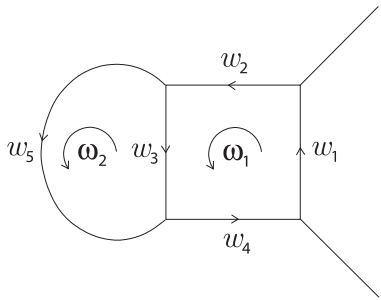,width=5cm} \caption{\small{A
two--loop graph, with loop winding numbers $\omega_1$, $\omega_2$.
The winding numbers $w_i$ for the propagators can be chosen
arbitrarily, as long as $ \omega_1 = \sum_{i=1}^4 w_i$ and
$\omega_2 = w_5 - w_3$.}} \label{FigLoopGraph}
\end{figure}

\item[{ii)}]  Fix external momenta $\mathbf{k}_{i}$ flowing into
the graph, and compute the diagram with the usual Feynman rules,
but with propagators given by the above prescription. The result
will depend on the explicit choice of representative $w_{i}\left(
\omega \right) $. Let us denote it by
\begin{equation*}
\Gamma _{w_{i}}\left( \mathbf{k}_{1},\cdots ,\mathbf{k}_{n}\right) \,.
\end{equation*}
The momenta $\mathbf{k}_{i}$ should be thought of as covering space momenta,
even though $\Gamma _{w_{i}}$ is not in general Lorentz invariant. However, since
invariance under $K_{+}$ is preserved, the above amplitude will
always contain a delta function
$2\pi\,\delta \left( \Sigma_{i}\,k_{i+}\right)\,$.

\item[{iii)}]  Average over the action of the group on the
external states as we did for the particle wave functions, by
considering the integral
\begin{equation}
\int ds_{1}\cdots ds_{n}\; e^{\,i\chi \left( s_{i}\right)}\;\Gamma_{w_{i}}
\left( \Omega ^{s_{1}}\mathbf{k}_{1},\cdots ,\Omega ^{s_{n}}\mathbf{k}_{n}\right) \,,
\label{amp1}
\end{equation}
with
\begin{equation*}
\chi \left( s_{i}\right) =
2\pi\tsum\nolimits_{i}s_{i}n_{i}-
\tsum\nolimits_{i}\varphi \left( \mathbf{k}_{i},s_{i}\right) \,.
\end{equation*}
The above integral overcounts the result, since invariance under the
orbifold action $\kappa $ implies that
\begin{equation*}
e^{\,i\chi \left( s_{i}\right) }\;\Gamma _{w_{i}}\left( \Omega ^{s_{i}}\mathbf{k}_{i}\right)
= e^{2\pi is\tsum\nolimits_{i}n_{i}}\;
e^{\,i\chi \left(s_{i}^{\prime }\right) }\;
\Gamma _{w_{i}}\left( \Omega ^{s_{i}^{\prime }}\mathbf{k}_{i}\right) \,,
\end{equation*}
where $s_{i}=s_{i}^{\prime }+s$. We may insert, in (\ref{amp1}), the
identity ``$1$''
\begin{equation*}
\left| \tsum\nolimits_{i}c_{i}\right|
\int ds\;\delta \left(\tsum\nolimits_{i}c_{i}s_{i}^{\prime }\right) \,,
\end{equation*}
where the constants $c_{i}$ can be chosen arbitrarily, as long as
$\tsum\nolimits_{i}c_{i}\neq 0$. Changing integration variables to the
$s_{i}^{\prime }$\thinspace , we can perform the integral over $s$ and obtain
a delta function $\delta \left( \tsum\nolimits_{i}n_{i}\right) $.
Restricting the integral over $s$ to a single action of the orbifold
generator, from $0$ to $1$, or following the prescription in (\ref{overcount}),
we can substitute this $\delta $--function with a Kronecher
symbol. Dropping the primes we arrive at the amplitude
\begin{equation}
\delta_{\,\Sigma\,n_{i}} \;\left| \tsum\nolimits_{i}c_{i}\right|\,
\int ds_{1}\cdots ds_{n}\;
\delta \left( \tsum\nolimits_{i}c_{i}s_{i}\right)\,
e^{\,i\chi \left( s_{i}\right) }\;
\Gamma_{w_{i}}\left( \Omega ^{s_{i}}\mathbf{k}_{i}\right) \,.
\label{basicamp}
\end{equation}
This expression depends only on the specific class $\omega \in
H^{1}\left( G\right)$ and it is the final result given a specific
choice of $\omega$ and of external states.

\item[{iv)}]  Finally sum over $\omega $. The term with $\omega
=0$ is singled out since it comes from the parent theory in
$\mathbb{M}^{3}$. In fact, we can choose $w_{i}=0$ for all
internal propagators and, in this case, $\Gamma _{w_{i}}\left(
\mathbf{k}_{i}\right) $ can be computed with the usual flat space
Feynman propagator. The only sign of the orbifold would then be in
the choice of external states. If $G$ is a tree--level graph, then
$H^{1}\left( G\right) =0$ and the only term in the sum comes from
the parent theory. This is usually called the inheritance
principle.
\end{enumerate}

To do computations, it is convenient to specialize the formula (\ref
{basicamp}) to the external states given by the functions $V_{\lambda
,p_{+},p_{-}}$ and $V_{p,p_{-}}$. Consider first the case when all $n$
external states have $p_{+}\neq 0$ and are labeled by $\lambda _{i}$, $%
p_{i+} $, $p_{i-}$. We must then compute the following integral (rescaling $%
c_{i}\rightarrow c_{i}p_{i+}$)
\begin{equation}
\delta_{\,\Sigma \,p_{i-}}\;
\left|\frac{\tsum\nolimits_{i}c_{i}p_{i+}}{\prod_{i}p_{i+}}\right|\,
\int dk_{1}\cdots dk_{n}\;
\delta \left( \tsum\nolimits_{i}c_{i}k_{i}\right)\,
e^{\,i\chi \left( k_{i}\right)}\;
\Gamma_{w_{i}}\left( \mathbf{k}_{i}\right) \,,  \label{A1}
\end{equation}
where the phase $\chi $ is given by
\begin{equation}
\chi \left( k_{i}\right) =\sum_{i}\frac{1}{2Ep_{i+}^{\,2}}\left[ \left(
2p_{i+}p_{i-}-\lambda _{i}\right) k_{i}-\frac{k_{i}^{\,3}}{3}\right] \,,
\label{phaseA1}
\end{equation}
and where
\begin{equation*}
k_{i+}=p_{i+}\,,\ \ \ \ \ \ \ \ \ \ \ \ \ \ \ \ \ \ \
k_{i-}=\frac{\lambda _{i}+k_{i}^{\,2}}{2p_{i+}}\,.
\end{equation*}
If, on the other hand, we have $n+1$ external states, one of which
has $p_{+}=0$ and therefore  is labeled by $p,p_{-}$, we may
choose the corresponding variable $s$ to vanish in
(\ref{basicamp}) and obtain
\begin{equation}
\delta_{\,p_{-}+\Sigma p_{i-}}\;
\left| \frac{1}{\prod_{i}p_{i+}}\right|\,
\int dk_{1}\cdots dk_{n}\;
e^{\,i\chi \left( k_{i}\right)}\;
\Gamma_{w_{i}}\left( \mathbf{k},\mathbf{k}_{i}\right)  \label{A2}
\end{equation}
where ${\bf k}=(0,0,p)$ and $i$ runs over the remaining $n$ external states.

\section{Propagation in the presence of CCC's\label{SectTwoPT}}

\begin{figure}[b!]
\centering\psfig{figure=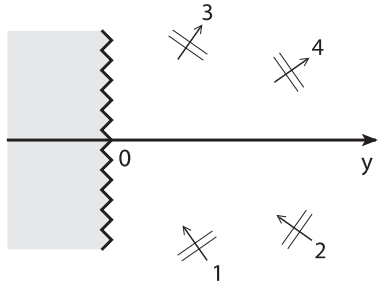,width=5cm}
\caption{\small{Scattering process in the orbifold geometry. The
dynamics is strongly coupled in the shaded region due to
gravitational interactions in the presence of CCC's, which naively
violate unitarity. We expect, on the other hand, to be able to
define particle states and a consistent S--matrix for discrete
values of the gravitational coupling constant $M^{-1}$.}}
\label{FigScattering}
\end{figure}

We have now the basic tools to address the main issue of this
paper, namely the analysis of particle propagation in the presence
of CCC's. We shall consider a scalar field with action
(\ref{action}), minimally coupled to gravity, and for simplicity
consider the massless case $m=0$. One expects that the existence
of CCC's will not allow a consistent definition of single particle
states and of an unitary $S$ matrix in the interacting theory.
This belief is supported by perturbative computations, since the
usual Cutkosky cutting rules are no longer valid, and unitarity
fails order by order in the coupling constant
\cite{Boulware:1992pm}. A simple example of this fact will be
described in section \ref{OnePF}. On the other hand, from the
duality arguments given in the introduction, one expects, for a
given value of the geometric parameter $E$, to be able to define
consistently a unitary $S$ matrix for the discrete values of the
Planck mass given by (\ref{Magic}). For those values of $M$, we
expect to be able to define particle states interacting strongly
with the gravitational field and scattering with unit probability,
as show pictorially in Figure \ref{FigScattering}.

\begin{figure}[t!]
\centering\psfig{figure=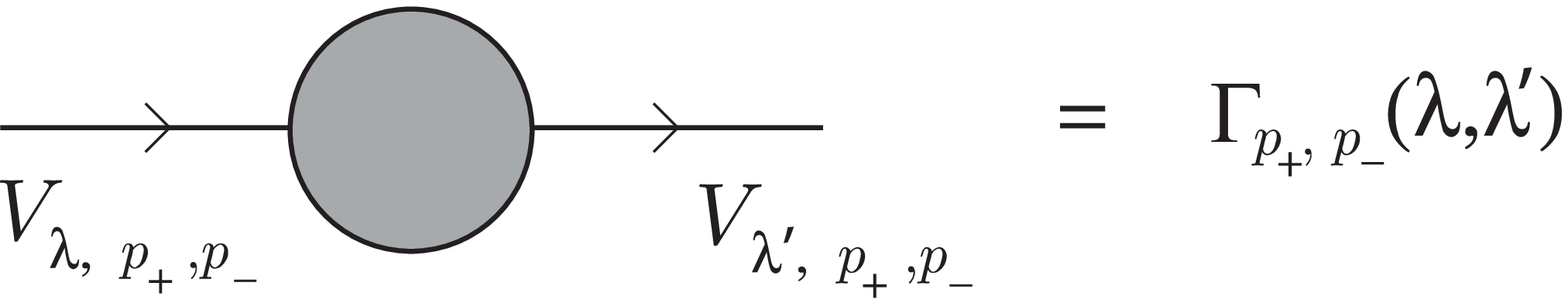,width=9cm} \caption{\small{The
correction to the free scalar propagator due to interactions. The
conserved momenta $p_\pm$ flow through the diagram, whereas
$\lambda,\lambda '$ are the off--shell mass squared of the
external legs. The shaded blob is computed using the Feynman rules
of section 3 and includes internal propagators winding the compact
$y^-$ direction.}} \label{FigTwoPTGraph}
\end{figure}

To investigate the possible restoration of unitarity, we shall
study the two--point function of the scalar field, which defines
single particle states. As shown in Figure \ref{FigTwoPTGraph},
this two--point function is determined by the conserved momenta
$p_{+},p_{-}$ flowing in the diagram, together with the off--shell
mass squared $\lambda ,\lambda ^{\prime }$ of the external legs.
We denote it by $\Gamma _{p_{+},p_{-}}\left( \lambda
,\lambda^{\prime }\right)$, so that the full propagator becomes
\begin{equation}
16\pi ^{2}E\,\left( \lambda +i\epsilon \right) \,\delta \left( \lambda
-\lambda ^{\prime }\right) +\Gamma _{p_{+},p_{-}}\left( \lambda ,\lambda
^{\prime }\right) \,.  \label{kernel}
\end{equation}
It is then possible to define consistently single particle states
whenever the above kernel has real eigenfunctions with vanishing
eigenvalue. This will certainly be the case if the effective
potential $\Gamma_{p_{+},p_{-}}$ determining the propagation of
the scalar field satisfies the reality condition
\begin{equation} \Gamma _{p_{+},p_{-}}^{\,\star }\left(
\lambda ,\lambda ^{\prime }\right) \,=\Gamma _{p_{+},p_{-}}\left(
\lambda ^{\prime },\lambda \right) \, . \label{reality}
\end{equation}
Note that, in quantum field theory, the potential
$\Gamma_{p_{+},p_{-}}$ can in principle have an imaginary part
coming from on--shell intermediate states. On the other hand, the
contribution to $\Gamma$ considered in this paper has no
contributions from intermediate on--shell lines and therefore any
deviation from the reality condition (\ref{reality}) would be a
sign of inconsistency of the theory. We shall comeback to this point at the
end of section 4.4.

\begin{figure}[t!]
\centering\psfig{figure=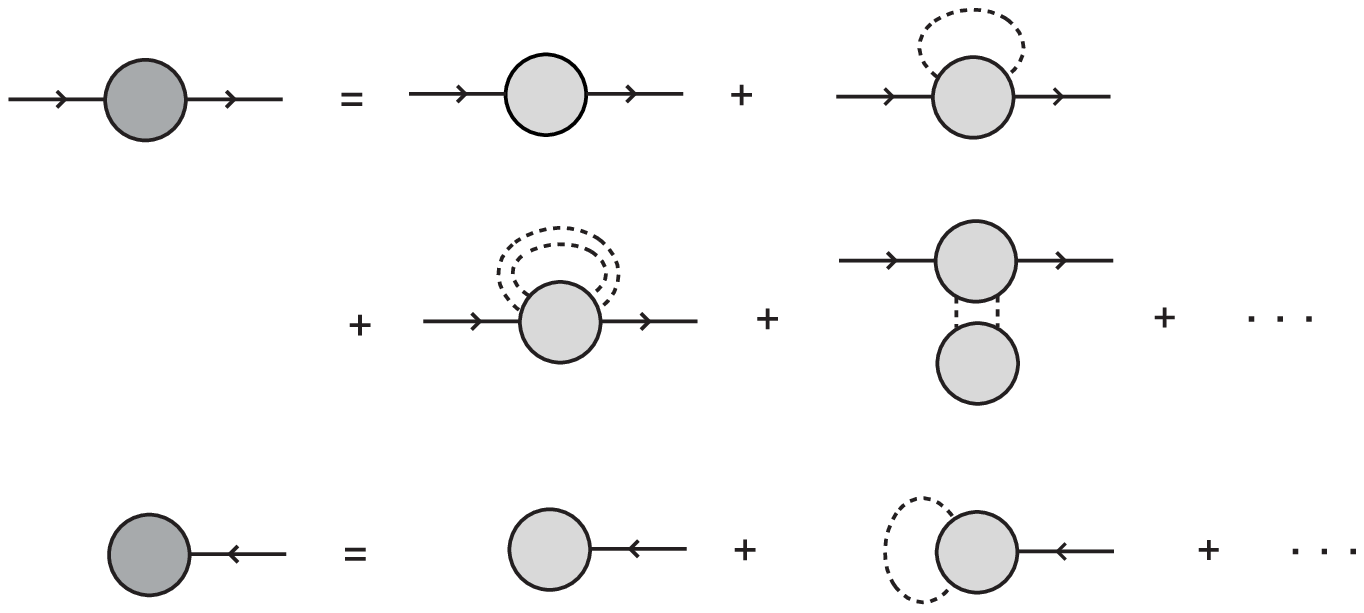,width=12cm}
\caption{\small{Complete scalar propagator $\Gamma
_{p_{+},p_{-}}\left( \lambda ^{\prime },\lambda \right)$ and
tadpole expanded in increasing number of internal propagators with
non--vanishing winding number, represented by dashed lines. The
remaining effective vertices, shown with light gray blobs, are
computed in the parent theory to all orders in the coupling
constant, with internal propagators with vanishing winding
number.}} \label{TwoPTEx}
\end{figure}

\begin{figure}
\centering\psfig{figure=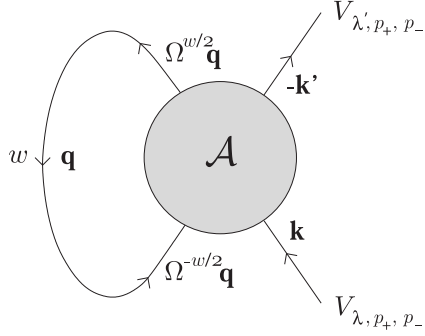,width=5.5cm}
\begin{picture}(0,0)(0,0)
\end{picture}
\caption{\small{Leading non--trivial contribution to the
two--point function $\Gamma_{p_+,p_-}(\lambda,\lambda')$, as shown
in Figure \ref{TwoPTEx}. The loop momentum has non--vanishing
winding number $w$, whereas the blob $\mathcal{A}$ represents the
four--point amplitude in the parent theory to all orders in the
coupling constants.}} \label{FigOneLoop}
\end{figure}

We will be able, in particular, to compute the first non--trivial
contribution to the full scalar propagator probing the non--causal
structure of space--time. As shown in Figure \ref{TwoPTEx}, we
consider an expansion of $\Gamma _{p_{+},p_{-}}\left( \lambda
^{\prime },\lambda \right)$ in increasing number of internal
propagators with non--vanishing winding number. We shall focus on
the leading non--trivial contribution arising from the graph
\ref{FigOneLoop}. The loop propagator will have non--vanishing
winding number $w$, whereas the bubble in the graph represents the
four--point interaction in the parent theory on $\mathbb{M}^{3}$
to all orders in the couplings. In the limit of $p_{+},\lambda
,\lambda ^{\prime }$ small, we shall see that, to compute the
graph, we will only need control over the parent four--point
amplitude in the eikonal kinematical regime, where resummation
techniques are known and where general arguments indicate that
interactions are dominated by graviton exchange. The form of the
eikonal amplitude in three dimensions is such that $\Gamma
_{p_{+},p_{-}}\left( \lambda ^{\prime },\lambda \right) $
satisfies (\ref{reality}) exactly at the values of the
gravitational coupling constant $M^{-1}$ given by (\ref{Magic}).

As a warm--up exercise, we compute, in the next subsection, the
scalar one--loop tadpole graph, so that the reader can get
acquainted with loop computations in the orbifold theory. This
simple computation already shows the breakdown of the cutting
rules.

\subsection{One--loop tadpole\label{OnePF}}

\begin{figure}
\centering\psfig{figure=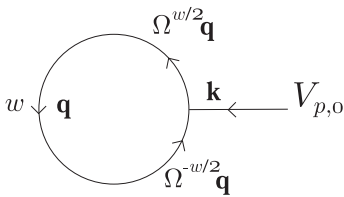,width=4.5cm}
\begin{picture}(0,0)(0,0)
\end{picture}
\caption{\small{The one--loop scalar tadpole. The on--shell
particle winding the compact $y^-$ direction contributes an
imaginary part to the tadpole, which is non--vanishing only for
$p<0$. In position space the tadpole diverges at the polarization
surfaces. }} \label{FigOneLoopTadpole}
\end{figure}

Let us explicitly compute the one--loop tadpole graph of Figure
\ref{FigOneLoopTadpole}, which contributes to the effective action with the term
\begin{equation}
\int_{\mathcal{F}}d^{3}\mathbf{x}\;\Phi(\mathbf{x}) \,\Gamma(\mathbf{x}) \,.
\label{eq2000}
\end{equation}
Conservation of $p_{\pm }$ forces the external leg to have quantum numbers $%
p_{\pm }=0$. Thus the one--loop tadpole can be written in momentum space as
\begin{equation*}
\Gamma \left( p\right) =\frac{1}{TE^{-1}}\int_{\mathcal{F}}d^{3}\mathbf{x}\;
V_{p,0}(\mathbf{x}) \,\Gamma(\mathbf{x}) \,,
\end{equation*}
where the wave functions $V_{p,0}$ were introduced in section \ref{wavefunctions}.
Changing to the $\mathbf{y}$--coordinates, it is easy to
see that $\Gamma(\mathbf{y})$ actually depends only on the
coordinate $y$, and that
\begin{equation}
\Gamma(\mathbf{y}) =\int \frac{dp}{2\pi }\;e^{-ipy}\;
\Gamma(p)\,\sqrt{\frac{ip}{2\pi E}}\,.
\label{Four1}
\end{equation}

To compute the function $\Gamma \left( p\right) $, we start with
the amplitude $\Gamma \left( \mathbf{k}\right) $ using Feynman
rule $ii)$ of last section
\begin{eqnarray*}
i\,\Gamma(\mathbf{k}) &=&\frac{ig}{2}\;\sum_{w\neq 0}\,
\int \frac{d^{3}q}{\left( 2\pi \right)^{3}}\;\frac{-i}{q^{2}-i\epsilon }\;
e^{\,\frac{i}{E}\left( wq_{-}+\frac{w^3}{24}\,q_{+}\right)}\times
\\
&&\times \left( 2\pi \right)^{3}\, \delta\left( k_{+}\right)\,
\delta\left(k_{-}-wq\right) \,\delta \left( k-wq_{+}\right) \\
&=&-2\pi\, \delta\left( k_{+}\right)\; \frac{ig}{2}\;\sum_{w\neq 0}\;
e^{\,\frac{i}{E}\frac{w^2k}{24}}\int \,\frac{dq_{-}}{2\pi i}\;
\frac{\,e^{\frac{i}{E}\,wq_{-}}}{2wq_{-}k-k_{-}^{\,2}+i\epsilon }\,,
\end{eqnarray*}
where the $w=0$ term is eliminated by renormalization of the parent theory.
The integral is non--vanishing only for $k<0$. Closing the $q_{-}$ counter
at infinity, the contribution of the pole, which is associated to the winding particle going
on--shell, gives
\begin{equation*}
\Gamma(\mathbf{k}) =2\pi\, \delta \left( k_{+}\right)\,
\theta(-k)\;\frac{g}{4\left| k\right|}\, \sum_{w\neq 0}\,
\frac{1}{\left|w\right| }\;
e^{\,\frac{i}{E}\left( \frac{w^{2}k}{24}+\frac{k_{-}^{\,2}}{2k}\right)}\,.
\end{equation*}
Finally, using (\ref{A2}), we conclude that $TE^{-1}\Gamma(p)=\Gamma(\mathbf{k})$,
with $\mathbf{k}=(0,0,p)$, so that the result is
\begin{equation}
\Gamma(p) = \theta(-p)\;\frac{gE}{4\left| p\right| }\,
\sum_{w\neq 0}\,\frac{1}{\left| w\right| }\;e^{\,\frac{i}{E}\frac{w^{2}p}{24}}\,.
\label{Four2}
\end{equation}

Let us note that, for the simple tadpole graph under consideration, one
could compute $\Gamma(\mathbf{x})$ directly in position space
\begin{equation*}
\Gamma(\mathbf{x}) = \,\frac{g}{2}\,\sum_{w\neq 0}\Delta
\left(\Omega ^{w}\mathbf{x},\mathbf{x}\right) \,.
\end{equation*}
Denoting with
\begin{equation*}
d_{w}(y) =\sqrt{\frac{2w^{2}}{E}\left( y-\frac{w^{2}}{24E}\right) +i\epsilon }\,,
\end{equation*}
the distance between the point $\mathbf{x}$ and its $w$--th image
as a function of the $y$--coordinate $y(\mathbf{x})$, we obtain
from the standard expression for the three--dimensional scalar
propagator
\begin{equation*}
\Gamma(\mathbf{y}) = \,\frac{g}{8\pi}\,\sum_{w\neq 0}\,
\frac{1}{d_{w}\left( y\right) }\,.
\end{equation*}
The reader can check, using (\ref{Four1}), that the Fourier transform of the
above expression yields (\ref{Four2}).

Clearly, for fixed $w$, the contribution to the one--loop tadpole diverges at the polarization
surface $y=w^{2}/\left( 24E\right)$ and becomes complex for
$y<w^{2}/\left(24E\right) $. Correspondingly, in momentum space, the contribution to
$\Gamma \left( p\right) $ is non--vanishing only for $p<0$ and comes uniquely
from on--shell particles running in the loop, which wind around the compact
circle and contribute with an imaginary part to the one--loop
tadpole.

In the computation of the graph in Figure \ref{FigOneLoopTadpole},
the three--vertex is probed at high energies, and therefore
interactions will drastically modify its behavior, even in the
parent theory. As for the two--point function, we should consider
the expansion in Figure \ref{TwoPTEx} and compute graph
\ref{FigOneLoopTadpole} with the complete parent three--point
coupling. This coupling is dominated by gravitational interactions
and can be resummed in some cases with eikonal techniques. On the
other hand, we shall not pursue this line any further and we shall
concentrate mostly on the most relevant computation of the
two--point function.

\subsection{Two--point function\label{TwoPTFunct}}

Let us consider now the quadratic term in the effective action, which is
composed of the free part in (\ref{action}), together with the contribution
from interactions given by
\begin{equation}
\frac{1}{2}\,\int_{\mathcal{F}}d^{3}\mathbf{x}\,d^{3}\mathbf{x}^{\prime}\;
\Phi\left(\mathbf{x}\right)\, \Phi\left(\mathbf{x}^{\prime}\right)
\,\Gamma \left( \mathbf{x},\mathbf{x}^{\prime }\right) .
\label{twoptfct}
\end{equation}
In momentum space the two--point function $\Gamma \left( \mathbf{x},\mathbf{x}%
^{\prime }\right) $ becomes
\begin{eqnarray*}
&&\hspace{-1in}\int_{\mathcal{F}}d^{3}\mathbf{x}\,d^{3}\mathbf{x}^{\prime}\;
V_{\lambda ,p_{+},p_{-}}\left( \mathbf{x}\right)\;
V_{\lambda ^{\prime},p_{+}^{\prime },p_{-}^{\prime }}\left( \mathbf{x}^{\prime }\right)
\;\Gamma \left( \mathbf{x},\mathbf{x}^{\prime }\right) = \\
&&=E^{-1}\;\delta_{p_{-}+p_{-}^{\prime }}\;2\pi\, \delta \left(
p_{+}+p_{+}^{\prime }\right) \,\Gamma _{p_{+},p_{-}}\left( \lambda ,\lambda
^{\prime }\right) \,,
\end{eqnarray*}
where $\Gamma _{p_{+},p_{-}}\left( \lambda ,\lambda ^{\prime }\right)
=\Gamma _{-p_{+},-p_{-}}\left( \lambda ^{\prime },\lambda \right) $. As for
the one--loop tadpole, the last equation can be explicitly computed using the
Feynman rules of section \ref{FeynRules}, in particular using equation (\ref
{A1}). It gives the two--point function written in the basis $V_{\lambda
,p_{+},p_{-}}$, with the conservation of $p_{\pm }$ momenta explicitly
stated. Reality of the interaction $\Gamma \left( \mathbf{x},\mathbf{x}%
^{\prime }\right) $ then reads
\begin{equation}
\Gamma _{p_{+},p_{-}}^{\,\star }\left( \lambda ,\lambda ^{\prime }\right)
=\Gamma _{p_{+},p_{-}}\left( \lambda ^{\prime },\lambda \right) =\Gamma
_{-p_{+},-p_{-}}\left( \lambda ,\lambda ^{\prime }\right) \,.
\label{realGamma}
\end{equation}
Using (\ref{A1}), we may write $\Gamma _{p_{+},p_{-}}\left( \lambda ,\lambda
^{\prime }\right) $ as follows
\begin{equation}
\Gamma _{p_{+},p_{-}}\left( \lambda ,\lambda ^{\prime }\right) =\,
\frac{2E}{T\left| p_{+}\right| }\int dk\;e^{\,i\chi \left( k\right)}\;
\Gamma \left(\mathbf{k},\mathbf{k}^{\prime }\right) \,,
\label{twopointfct}
\end{equation}
where the phase $\chi(k)$ is explicitly given by
\begin{equation*}
\chi(k) =\frac{1}{Ep_{+}^{\,2}}\left[ \left( 2p_{+}p_{-}-\frac{%
\lambda +\lambda ^{\prime }}{2}\right) k-\frac{k^{3}}{3}\right] \,.
\end{equation*}
The amplitude $\Gamma \left( \mathbf{k},\mathbf{k}^{\prime}\right)$
is computed using Feynman rule $ii)$ of section
\ref{FeynRules}, with external momenta given by
\begin{equation*}
\mathbf{k}=\left( p_{+},\frac{\lambda +k^{2}}{2p_{+}},k\right) \,,\,\ \ \ \
\ \ \ \ \ \ \mathbf{k}^{\prime }=\left( -p_{+},-\frac{\lambda ^{\prime
}+k^{2}}{2p_{+}},k\right) \,.
\end{equation*}

We now focus our attention on external states in the two--point
function such that the integral (\ref{twopointfct}) can be computed using
the saddle point approximation. This requires that $4p_{+}p_{-}>\lambda
+\lambda ^{\prime }$, in order to have saddle points on the real $k$ axis at
$k=\pm p$, with
\begin{equation*}
p=\sqrt{2p_{+}p_{-}-\frac{\lambda +\lambda ^{\prime }}{2}}\,.
\end{equation*}
Then, the gaussian approximation to (\ref{twopointfct}) is valid provided
that
\begin{equation}
Ep_{+}^{\,2}\ll p^{3}\,.
\label{saddleOK}
\end{equation}
In this situation, the two--point function becomes
\begin{equation*}
\Gamma _{p_{+},p_{-}}\left( \lambda ,\lambda ^{\prime }\right) \simeq
\frac{E}{T}\,\sqrt{\frac{4\pi E}{ip}}\;e^{\,\frac{2i}{3}\frac{p^{3}}{Ep_{+}^{\,2}}}\;
\Gamma \left( \mathbf{k},\mathbf{k}^{\prime }\right) +\left(p\leftrightarrow -p\right) \,,
\end{equation*}
where the external momenta $\mathbf{k},\,\mathbf{k}^{\prime }$ are fixed,
at the saddle point $k=p$, to be
\begin{equation*}
\mathbf{k}=\left( p_{+},p_{-}+\frac{\lambda -\lambda ^{\prime }}{4p_{+}},p\right)
\,,\,\ \ \ \ \ \ \ \ \ \ \ \mathbf{k}^{\prime }=\left(
-p_{+},-p_{-}+\frac{\lambda -\lambda ^{\prime }}{4p_{+}},p\right) .
\end{equation*}

We consider in what follows the first non--trivial two--point
graph probing the non--causal structure of space--time, which has
a single scalar propagator with non--vanishing winding number $w$.
This graph is shown in Figure \ref{FigOneLoop}, where the bubble
represents the four--point amplitude in the parent theory which we
denote by
\begin{equation*}
\left( 2\pi \right)^{3}\;\delta^{3}
\left(\Omega^{-w/2}\mathbf{q}+\mathbf{k}+\mathbf{k}^{\prime}-\Omega^{w/2}\mathbf{q}\right)\,
\mathcal{A}\left(\Omega ^{-w/2}\mathbf{q,k,k}^{\prime },-\Omega ^{w/2}\mathbf{q}\right) \,.
\end{equation*}
Since $\left(\Omega^{-w/2}\mathbf{q}\right)^{2}=\left( \Omega
^{w/2}\mathbf{q}\right)^{2}$, the amplitude $\mathcal{A}$ depends
only on five kinematical invariants. Moreover, since $\lambda
=-\mathbf{k}^{2}$ and $\lambda ^{\prime }=-\mathbf{k}^{\prime 2}$
are kept fixed, $\mathcal{A}$ depends only on the Mandelstam
invariants
\begin{eqnarray*}
s &=&-\left( \mathbf{k+}\Omega ^{-w/2}\mathbf{q}\right) ^{2}\,, \\
t &=&-\left( \mathbf{k+k}^{\prime }\right) ^{2}\,, \\
u &=&-\left( \mathbf{k-}\Omega ^{w/2}\mathbf{q}\right) ^{2}\,.
\end{eqnarray*}
Finally, again because $\left( \Omega ^{-w/2}\mathbf{q}\right) ^{2}=\left(
\Omega ^{w/2}\mathbf{q}\right) ^{2}$, the amplitude $\mathcal{A}\left(
s,t,u\right) $ is symmetric under interchange of $s\leftrightarrow u$ and of
$\lambda \leftrightarrow \lambda ^{\prime }\,$.

For simplicity, we consider first the case of on--shell external states with
$\lambda =\lambda ^{\prime }=0$. In this case, in order to satisfy (\ref
{saddleOK}), we will take the limit of small $p_{+}$, since $p_{-}$ and $E$
are fixed. Then, the contribution to $\Gamma \left( \mathbf{k},\mathbf{k}%
^{\prime }\right) $ of the graph with winding number $w$ is
\begin{eqnarray}
&&\Gamma _{w}\left( \mathbf{k,k}^{\prime }\right) =\,\frac{1}{2}
\int \frac{d^{3}\mathbf{q}}{\left( 2\pi \right)^{3}}\;\frac{i}{-\mathbf{q}^{2}+i\epsilon}\;
e^{\,\frac{i}{E}\left( wq_{-}+\frac{w^3}{24}\,q_{+}\right)}\times
\label{saddleintegral} \\
&&\hspace{0.6cm}\times\left(2\pi\right)^{3}\;\delta^{3}
\left( \Omega^{-w/2}\mathbf{q} + \mathbf{k}+\mathbf{k}^{\prime}-\Omega^{w/2}\mathbf{q}\right)\,
\mathcal{A}\left( \Omega ^{-w/2}\mathbf{q,k,k}^{\prime },-\Omega ^{w/2}
\mathbf{q}\right)  \notag \\
&&\hspace{0.3cm}=-\frac{T}{8}\;e^{\,\frac{i}{E}\frac{w^{2}p}{12}}\int \frac{dq_{-}}{2\pi i}\;
\frac{1}{pwq_{-}+i\epsilon }\;e^{\,\frac{i}{E}wq_{-}}\;\mathcal{A}
\left(\Omega ^{-w/2}\mathbf{q,k,k}^{\prime },-\Omega ^{w/2}\mathbf{q}\right) \,.
\notag
\end{eqnarray}
Two of the $\delta $--functions fix $q_{+}=2p/w$ and $q=0$, while the
remaining $\delta $--function gives the overall factor $T$. In the last
line, we therefore have that
\begin{eqnarray*}
\Omega ^{-w/2}\mathbf{q} &=&\left( \frac{2p}{w},q_{-}+\frac{wp}{4},-p\right)
\,, \\
-\Omega ^{w/2}\mathbf{q} &=&-\left( \frac{2p}{w},q_{-}+\frac{wp}{4},p\right)
\,.
\end{eqnarray*}
The corresponding Mandelstam invariants can be readily computed
\begin{eqnarray}
s &=&2\left( \frac{2p}{w}+p_{+}\right) \left( q_{-}+\frac{wp}{4}%
+p_{-}\right) \simeq \frac{4p}{w}\left( q_{-}+p_{-}\right) \,,  \notag \\
t &=&-4p^{2}\,,  \label{sut} \\
u &=&2\left( \frac{2p}{w}-p_{+}\right) \left( q_{-}+\frac{wp}{4}%
-p_{-}\right) \simeq \frac{4p}{w}\left( q_{-}-p_{-}\right) \,,  \notag
\end{eqnarray}
where we have expanded $s$ and $u$ to  leading order in $p\sim \sqrt{%
\left| p_{+}\right| }$, since we are working in the limit $p_{+}\rightarrow
0 $. More precisely, we will be working under the assumption that
\begin{equation}
p\ll \frac{E}{\left| w\right| }\,,  \label{eikOK}
\end{equation}
which implies (\ref{saddleOK})\ but is more stringent for $\left| w\right|
\gg 1$.

The integral (\ref{saddleintegral}) reads
\begin{equation*}
\Gamma _{w}\left( \mathbf{k,k}^{\prime }\right) \simeq -\frac{T}{8}\;
e^{\,\frac{i}{E}\frac{w^{2}p}{12}}\int \frac{dq_{-}}{2\pi i}\;\frac{1}{pwq_{-}+i\epsilon }\;
e^{\,\frac{i}{E}wq_{-}}\;\mathcal{A}\left( s,t,u\right) \,,
\end{equation*}
where the Mandelstam invariants are given above. Notice that this
integral is invariant under $w\rightarrow -w$. This can be seen by changing
integration variable to $-q_{-}$ and by using $s\leftrightarrow u$
invariance of $\mathcal{A}$. For the same reasons, changing $p\rightarrow -p$
has the unique effect of changing the phases to their complex conjugates
$e^{-\frac{i}{E}wq_{-}}$ and $e^{-\frac{i}{E}\frac{w^{2}p}{12}}$.
Using (\ref{twopointfct}), we conclude that the two--point amplitude can be written as
\begin{equation*}
\Gamma_{p_{+},p_{-}}\left( 0,0\right) \simeq \sum_{w\neq 0}
\left(c_{w}\,\Gamma _{w}^{+}+c_{w}^{\,\star}\,\Gamma_{w}^{-}\right) \,,
\end{equation*}
where
\begin{equation}
\Gamma _{w}^{\pm }=\int \frac{ds}{2\pi i}\;
\frac{e^{\pm \frac{i}{4}\frac{w^{2}}{pE}s}}{\left| w\right|s-4pp_{-}+i\epsilon}\;
\mathcal{A}\left(s,t=-4p^{2},u=s-\frac{8pp_{-}}{\left| w\right| }\right) \,  \label{AmpInt}
\end{equation}
and the constant $c_{w}$ is given by
\begin{equation}
c_{w}=-\frac{E}{8p}\,\sqrt{\frac{4\pi E}{ip}}\;
e^{\,\frac{i}{E}\left( \frac{2p^{3}}{3p_{+}^{\,2}}+\frac{w^{2}p}{12}\right)}\,.
\label{constFinal}
\end{equation}
The reality condition (\ref{realGamma}) implies that
$\Gamma_{p_{+},p_{-}}(0,0)$ is real, which in turn implies that
$\left( \sum_{w}c_{w}\,\Gamma _{w}^{+}\right) ^{\star }=
\sum_{w}c_{w}^{\,\star}\,\Gamma _{w}^{-}$. On physical grounds, we expect the stronger condition
\begin{equation}
\left( \Gamma _{w}^{+}\right) ^{\star }=\Gamma _{w}^{-}  \label{singleReal}
\end{equation}
to hold for each value of $w$. This can be understood, for example, from the
behavior of the one--loop tadpole in position space. For a given value of $w$,
the contribution to the tadpole diverges and acquires an imaginary
part at the $w$--th polarization surface. Therefore, in position space, this
singular behavior occurs at widely separated positions for different values
of the winding number $w$ and we expect these pathologies to be cured for
each $w$.

We close this section with an important kinematic consideration.
When the winding particle is on--shell, then $s=-u=$ $%
4pp_{-}/\left| w\right| $. Since $|t|\sim p^{2}$, the parent
amplitude is evaluated in the eikonal regime
\begin{equation*}
|s|,|u|\gg |t|\,.
\end{equation*}
Moreover, this regime continues to be valid throughout the whole integration
region in $s$. In fact, for the non--generic case where $\left| s\right|
\lesssim \left| t\right| $, the other invariant $u$ is of order $%
pp_{-}/\left| w\right| $ and still satisfies $u\gg t$. A similar remark
applies when $\left| u\right| \lesssim \left| t\right| $. In the following section
we do a detour on the eikonal approximation to the scattering of a scalar field
in three dimensions. The results will be used in the computation of the orbifold
two--point amplitude.

\subsection{Eikonal approximation\label{EikSect}}

\begin{figure}
\centering\psfig{figure=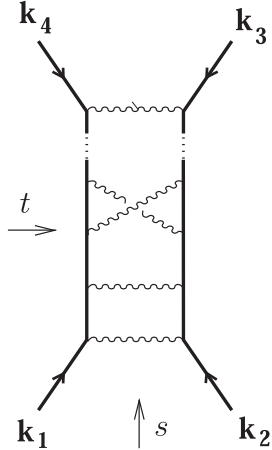,width=3.5cm}
\caption{\small{Generalized ladder diagram contributing to the
scattering amplitude $\mathcal{A}$ of scalar particles in the
eikonal regime $s\gg |t|$. The exchanged particle can have in
general spin $j$ but, in the kinematical regime of interest, the
gravitational interaction $j=2$ will dominate.}} \label{FigLadder}
\end{figure}

Let us recall first the $s$--channel partial wave decomposition for the
on--shell scattering amplitude of massless scalars in three dimensions
\begin{equation*}
1+i\mathcal{A}=4\sqrt{s}\,\sum_{n}\,e^{in\theta }\;e^{2i\delta _{n}\left(
s\right) }\,.
\end{equation*}
The scattering angle $\theta $ is given by $\sin ^{2}\left( \theta /2\right)
=-t/s$ and the phase shifts satisfy $\delta _{n}=\delta _{-n}$\ . Unitarity
requires that\ \ $\func{Im}\delta _{n}\geq 0$. In the eikonal limit $s\simeq
-u\gg -t$, the scattering angle is given by $\theta \simeq 2\sqrt{-t/s}$ and
one may replace the sum over partial waves with an integral over the impact
parameter $x=2n/\sqrt{s}$. One then obtains the eikonal expression
\begin{equation}
1+i\mathcal{A}\,\simeq\, 2s\int dx\;e^{ix\sqrt{-t}}\;e^{2i\delta \left(
s,x\right) }\,.  \label{eik1}
\end{equation}
This representation can be derived by studying the $s\gg |t|$
limit of the generalized ladder graphs shown in Figure
\ref{FigLadder}. It turns out that the loop expansion matches the
expansion in powers of $\delta$
\cite{'tHooft:1987rb,Levy:1969cr,Amati:1987wq,Kabat:1992tb}, so
that the phase shift is given, in terms of the leading tree--level
interaction $\mathcal{A} _{\mathrm{tree}}\left( s,t\right) $, by
the simple Fourier transform
\begin{equation*}
\mathcal{A}_{\mathrm{tree}}\,\simeq \,4s\int dx\;e^{ix\sqrt{-t}}\;
\delta(s,x) \,.
\end{equation*}

\begin{figure}
\centering\psfig{figure=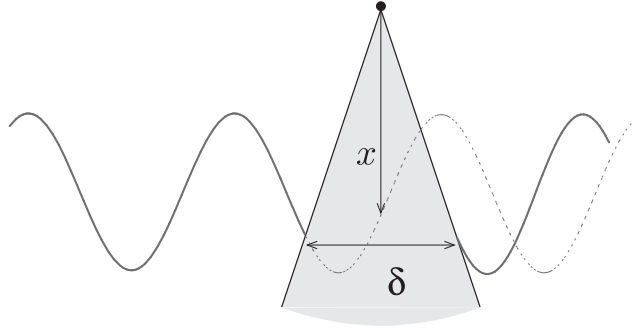,width=12cm}
\caption{\small{Phase shift for a scattering process in the
gravitational background created by the target. In three
dimensions, the background geometry is a conical space with
deficit angle proportional to $\sqrt{s}/M$. The phase shift
$\delta$ is proportional to the impact parameter $x$ and is
negative.}} \label{FigPhaseShift}
\end{figure}

Consider now scattering due to the exchange of a spin $j$ massless
particle. Then $\mathcal{A}_{\mathrm{tree}}\simeq
-4M^{3-2j}\,s^{j}/t$, where we are implicitly assuming that the
coupling constants are of order one in Planck units\footnote{We
assume that all couplings are of the same order in Planck units
mostly for notational simplicity. In case of a large ratios
between the couplings, the gravitational interaction will still
dominate in the regimes of interest, as explained in the rest of
this section, but the specific bounds will have to be modified
accordingly. Moreover, the most relevant examples coming from
compactifications of supergravity theories do have a single
gravitational coupling.}. The phase shift is then given by
\begin{equation}
2\,\delta(s,x) \simeq -\left( s/M^{2}\right) ^{j-1}M\left|
x\right| \,.  \label{phaseshift}
\end{equation}
It is linear in the impact parameter $\left| x\right| $ and
negative. These facts are easily understood in the $j=2$ case of
graviton exchange. In this case, following the work of 't Hooft
\cite{'tHooft:1987rb}, we can think of the scattered particle as
moving in the conical geometry created by the target. The
qualitative features of (\ref{phaseshift}) can then be immediately
understood from Figure \ref{FigPhaseShift}. The full eikonal
amplitude for spin--$j$ exchange finally reads
\cite{Deser:1988qn,'tHooft:1988yr,Deser:1993wt}
\begin{equation}
1+i\mathcal{A}\simeq -4iM\,\frac{\left( s/M^{2}\right) ^{j}}{\left(
t/M^{2}\right) +\left( s/M^{2}\right) ^{2j-2}-i\epsilon }\,.
\label{eikAmplitude}
\end{equation}
This result has been quoted in the literature only for the case
$j=2$, although its derivation easily extends to the case of
general $j$, as shown in the above derivation.

The eikonal amplitude has a pole at real values of the kinematical
invariants given by $-t=M^{2}\left( s/M^{2}\right) ^{2j-2}$. We
shall call this pole the 't Hooft pole, in analogy with those
discussed in  \cite{'tHooft:1987rb} for graviton exchange in four
dimensions. At the pole, the $i\epsilon$ prescription is obtained
by requiring convergence of the integral (\ref {eik1}) at large
values of $\left| x\right| $. One can physically understand this
prescription by first noting that, at vanishing coupling and
phase--shift, (\ref{eik1}) gives the $S$--matrix element
\begin{equation*}
4\pi s\;\delta \left( \sqrt{-t}\right) .
\end{equation*}
This amplitude corresponds to free propagation of particles, with no
interaction. Using the fact that $\func{Im}(x-i\epsilon )^{-1}=\pi \delta
\left( x\right) $ in the amplitude (\ref{eikAmplitude}), we then see that
the $\delta $--function contribution in the free theory is replaced, in the
interacting theory, by
\begin{equation*}
2\pi s\;\delta \left( \sqrt{-t}-M\left( s/M^{2}\right) ^{j-1}\right) +2\pi
s\;\delta \left( \sqrt{-t}+M\left( s/M^{2}\right) ^{j-1}\right) \,.
\end{equation*}
We conclude that a specific scattering angle, dependent on the energy of the
process, is singled out by the eikonal amplitude. Notice that the pole is in
the physical eikonal region $s\gg -t$ whenever $\left( s/M^{2}\right)
^{2j-3}\ll 1$. This shows that gravitons behave quite differently from lower
spin particles. In fact, for $j=2$, the eikonal approximation is reliable
around the 't Hooft pole for center of mass energies well below the Planck
mass.

We may now discuss which interaction dominates in the kinematical regime of
interest in (\ref{AmpInt}). More specifically, we will concentrate on the
cases $j=0,1,2$, which arise in standard compactifications of supergravity
theories. We will compare the relative importance of the various
contributions when the propagator in (\ref{AmpInt}) is on--shell, that is at $%
s=4pp_{-}/\left| w\right| $ and $t=-4p^{2}$. We are only
interested in an order of magnitude estimate, dropping factors of
order unity. Assume that the Ka\l u\.{z}a--Klein charge
$n=p_{-}/\left( 2\pi E\right) $ is of order unity and that
\begin{equation*}
E\sim \frac{M}{J}\,,
\end{equation*}
with $J\gtrsim 1$, which is consistent with
the quantization condition (\ref{Magic}). The kinematical invariants are
then given by
\begin{equation*}
s\sim \frac{pE}{w}\sim \frac{pM}{wJ}\,,\ \ \ \ \ \ \ \ \ \ \ \ \ \
\ |t|\sim p^{2}\,,
\end{equation*}
and the basic requirement $s\gg \left| t\right| $ for the eikonal
approximation to be valid is
\begin{equation*}
p\ll \frac{M}{wJ}\,.
\end{equation*}
When this condition is satisfied, the saddle point approximation (\ref
{saddleOK}) in the previous subsection is also justified.

Next we analyze the behavior of the eikonal amplitude (\ref{eikAmplitude})
for different values of $j$. For $j=0,1$, the denominator in (\ref
{eikAmplitude}) is dominated by the second term, whereas for $j=2$ the first
term dominates. We therefore arrive at the following estimates for the
amplitudes
\begin{eqnarray*}
M\left( s/M^{2}\right) ^{2-j} &\sim &M\left( \frac{p}{MwJ}\right) ^{2-j}\;\ll\;
\frac{M}{\left( wJ\right) ^{4-2j}}\;,\ \ \ \ \ \ \ \ \ \
\left(j=0,1\right) \\
\frac{1}{M}\,\frac{s^{2}}{t} &\sim &\frac{M}{\left( wJ\right) ^{2}}\;.\ \ \ \
\ \ \ \ \ \ \ \ \ \ \ \ \ \ \ \ \ \ \ \ \ \ \ \ \ \ \ \ \ \ \ \ \ \ \ \ \ \ \
\left( j=2\right)
\end{eqnarray*}
Thus, the graviton interaction dominates in the kinematical region of
interest. Therefore, from now on, we will only consider the case $j=2$.

Up to this point, we have discussed the amplitude $\mathcal{A}$
on--shell, with $s\simeq -u$. On the other hand, as it is clear
from (\ref{AmpInt}), one needs to understand the off--shell
extension of the amplitude (\ref {eikAmplitude}) for generic
values of $s,u$. We also allow in principle for non--vanishing
$\lambda ,\lambda ^{\prime }$, small compared to $s,u$. In what
follows, we shall indicate explicitly only the dependence on the
``large'' variables $s,u$, leaving implicit the dependence on the
``small'' ones $t,\lambda ,\lambda ^{\prime }$. The tree level
result is simple to compute and gives
\begin{equation}
\mathcal{A}_{\mathrm{tree}}\simeq -\frac{\left( s-u\right) ^{2}}{Mt}\,.
\label{tree}
\end{equation}
To understand the off--shell amplitude to all orders, we first rewrite the
on--shell result (\ref{eikAmplitude}) as follows
\begin{equation}
\mathcal{A}_{+}\simeq -\frac{\left( s-u\right) ^{2}}{2\sqrt{-t}}\left( \frac{%
1}{s-M\sqrt{-t}-i\epsilon }+\frac{1}{u-M\sqrt{-t}-i\epsilon }\right) \,.
\label{ampSplit}
\end{equation}
We neglect from now on the free propagation term $4\pi s\delta \left( \sqrt{%
-t}\right) $ in the $S$--matrix element (\ref{eikAmplitude}) since it does
not contribute to the orbifold amplitude (\ref{AmpInt}), which is computed
at fixed non--vanishing momentum transfer. One may interpret (\ref{ampSplit})
as the exchange, in the $s$ and $u$ channels, of an effective particle of
mass squared $M\sqrt{-t}$. Note, however, that the $i\epsilon $
prescription, which arises from the eikonal result, is opposite to the usual
one. Alternatively, recalling that $s\simeq -u$ on--shell, one rewrites
(\ref{ampSplit}) as
\begin{equation*}
\mathcal{A}_{-}\simeq \frac{\left( s-u\right) ^{2}}{2\sqrt{-t}}
\left(\frac{1}{s+M\sqrt{-t}+i\epsilon}+\frac{1}{u+M\sqrt{-t}+i\epsilon}\right)\,,
\end{equation*}
which can be viewed as resulting from the exchange of a tachyonic effective
particle of mass squared $-M\sqrt{-t}$, with the usual $i\epsilon $
prescription.

Given the above observations, we shall assume that the off--shell
extension of the eikonal amplitude has poles determined by
effective particles with mass squared $\pm M\sqrt{-t}$. The
amplitude $\mathcal{A}\left( s,u\right) $ therefore has poles
placed at
\begin{equation*}
s=M\sqrt{-t}+i\epsilon \,,\ \ \ \ \ \ \ \ \ \ \ \ \ \ \ \ \ \ \ \
u=M\sqrt{-t}+i\epsilon \,,
\end{equation*}
and at
\begin{equation*}
s=-M\sqrt{-t}-i\epsilon \,,\ \ \ \ \ \ \ \ \ \ \ \ \ \ \ \ \ \ \ \
u=-M\sqrt{-t}-i\epsilon \,.
\end{equation*}
We denote the residues of the amplitude $\mathcal{A}\left( s,u\right) $ by
\begin{eqnarray*}
\mathrm{Res}_{s}\text{ }\mathcal{A}\left( s=\pm M\sqrt{-t},u=s-\sigma
\right) \, &=&f_{\pm }(\sigma) \,, \\
\mathrm{Res}_{u}\text{ }\mathcal{A}\left( s=u+\sigma ,u=\pm M\sqrt{-t}
\right) &=&f_{\pm }(-\sigma) \,,
\end{eqnarray*}
where the real functions $f_{\pm }$ are the same at the $s$ and
$u$ poles because of $s,u$ symmetry of the amplitude. For later
convenience, we expressed  the above functions $f_{\pm }$ in terms
of $\sigma =s-u$.

It is important to stress that, in order to compute the orbifold two--point
function and to check the quantization condition (\ref{Magic}), we shall only assume
that the poles of the eikonal amplitude are placed at $s,u=\pm M\sqrt{-t}$, as for the on--shell
case. The $i\epsilon $ prescription, on the other hand, is fixed by the
on--shell computation. Finally, the reality of the functions $f_{\pm }$
follows from the fact that, in the absence of discontinuities, field
theoretic amplitudes are real.

A simple example of an off--shell extension of the eikonal amplitude
(\ref{eikAmplitude}), satisfying the above requirements, is
\begin{equation}
a_{+}\mathcal{A}_{+}+a_{-}\mathcal{A}_{-}\,,  \label{example}
\end{equation}
with $a_{\pm }$ constant and $\mathcal{A}_{\pm}$ as defined above.
To match the off--shell tree--level result (\ref{tree}) and the
on--shell eikonal amplitude we must have that
\begin{equation}
a_{+}+a_{-}=1.  \label{exCondition}
\end{equation}

\subsection{Quantization condition from unitarity\label{QCFU}}

\begin{figure}
\centering\psfig{figure=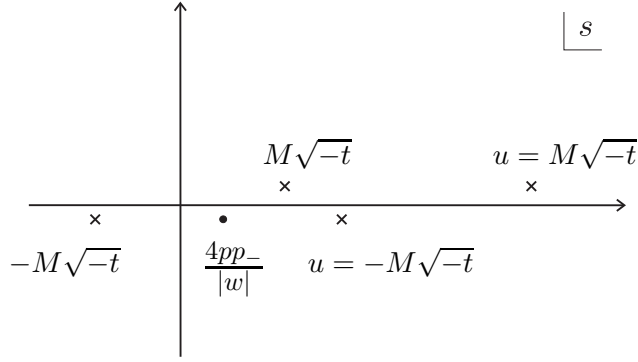,width=8cm}
\begin{picture}(0,0)(0,0)
\put(-238,33){{$-M\sqrt{-t}$}}
\put(-165,31){\Large{$\frac{4pp_-}{{|w|}}$}}
\put(-125,33){{$u=-M\sqrt{-t}$}} \put(-142,74){{$M\sqrt{-t}$}}
\put(-55,74){{$u=M\sqrt{-t}$}}
\end{picture}
\caption{\small{Poles of the integrand in equation
(\ref{finalint}) in the $s$--plane. The pole denoted with a dot
comes from the winding propagator, whereas the poles marked with a
cross come from the eikonal amplitude $\mathcal{A}$. Recall that
$s=u+\sigma$, with $\sigma=8pp_-/|w|$.}} \label{FigPoleStruct}
\end{figure}

We are now in position to finish the computation of the two--point
function for a massless scalar field in the orbifold geometry. As shown in
sections \ref{TwoPTFunct} and \ref{EikSect}, this amounts to evaluating the
integral
\begin{equation}
\Gamma _{w}^{\pm }=\int \frac{ds}{2\pi i}\;
\frac{e^{\,\pm \frac{i}{4}\frac{w^{2}}{pE}s}}{\left| w\right| s-4pp_{-}+i\epsilon }\;
\mathcal{A}\left(s,t=-4p^{2},u=s-\frac{8pp_{-}}{\left| w\right| }\right)\,,
\label{finalint}
\end{equation}
where the amplitude $\mathcal{A}$ is dominated
by graviton exchange in the eikonal regime. The integrand has a pole at
$s=4pp_{-}/w-i\epsilon $, coming from the winding propagator, and poles at
$s=\pm M\sqrt{-t}\pm i\epsilon $ and at $u=\pm M\sqrt{-t}\pm i\epsilon $, from
the eikonal amplitude $\mathcal{A}$. The pole structure of the integrand is
shown in Figure \ref{FigPoleStruct}.

To compute $\Gamma _{w}^{+}$ one closes the $s$--contour in the upper half
plane, so that the integral is determined by the poles of $\mathcal{A}$ with
positive imaginary part at $s,u=M\sqrt{-t}+i\epsilon =2pM+i\epsilon $. The
phase in (\ref{finalint}) is given at both poles by $e^{\,i\frac{w^{2}M}{2E}}$,
since $p_{-}$ is quantized in units of $2\pi E$. One then obtains
\begin{equation*}
\Gamma _{w}^{+}=\frac{1}{\left| w\right| }\;e^{\,i\frac{w^{2}M}{2E}}F_{+}\,,
\end{equation*}
where $F_{+}$ is given by the real constant
\begin{equation*}
F_+=\frac{2f_{+}(\sigma)}{4pM-\sigma }+\frac{2f_{+}(-\sigma)}{4pM+\sigma}
\end{equation*}
and where $\sigma =s-u=8pp_{-}/\left| w\right| $. Similarly,
$\Gamma _{w}^{-}$ is determined by the poles in the lower half
$s$--plane, with the result
\begin{equation*}
\Gamma _{w}^{-}=\frac{1}{\left| w\right| }\left( e^{\,i\frac{w^{2}M}{2E}}F_--
\mathcal{A}_{\mathrm{on-shell}}\right) \,,
\end{equation*}
where now
\begin{equation*}
F_-=\frac{2f_{-}(\sigma)}{4pM+\sigma }+\frac{2f_{-}(-\sigma)}{4pM-\sigma }\,.
\end{equation*}
The on--shell amplitude
\begin{equation}
\mathcal{A}_{\mathrm{on-shell}}=-4M\,\frac{4p_{-}^{2}}{4p_{-}^{2}-w^{2}M^{2}}\,
=\frac{4M\sigma ^{2}}{16M^{2}p^{2}-\sigma ^{2}}
\label{onShellA}
\end{equation}
is computed at the pole of the winding propagator.

Finally, we can investigate the implication of the reality condition (\ref
{singleReal}), which now reads
\begin{equation}
-e^{\,i\frac{w^{2}M}{2E}}F_++e^{-i\frac{w^{2}M}{2E}}F_-=
\mathcal{A}_{\mathrm{on-shell}}\,.  \label{MagicEQ}
\end{equation}
We consider the equation above for different values of $w$. In
order for the inequality (\ref{eikOK}) to be valid for all $w$, we
keep $p_{-}$ fixed and tune $p_{+}$ so that $p=\Lambda /\left|
w\right|\,$, for a fixed energy scale $\Lambda \ll E$. As
$w\rightarrow \infty $, we then have that
$\sigma =8\Lambda p_{-}/w^{2}\rightarrow 0 $. Note that $F_+$, $F_-$ and $\mathcal{A}_{%
\mathrm{on-shell}}$ are all analytic functions of $\sigma $, with
at most poles and branch cuts. This is clear for (\ref{onShellA}),
and it follows from general properties of analyticity of Feynman
amplitudes for the functions $f_{\pm }$, and therefore for
$F_{\pm}$. As $w\rightarrow \infty $, the phase
$e^{\,i\frac{w^{2}M}{2E}}$ oscillates unless
$e^{\,i\frac{M}{2E}}=1$, whereas the functions $F_+$, $F_-$ and
$\mathcal{A}_{\mathrm{on-shell}}$ have a regular behavior for
$\sigma \rightarrow 0$. Under mild regularity assumptions, we
show\footnote{We thank M. Cornalba for suggesting this argument.}
in appendix B that, in order for (\ref{MagicEQ}) to be satisfied
for all values of $w$, we must have that $e^{\,i\frac{M}{2E}}=1$
and therefore that
\begin{equation*}
\frac{M}{2E}\,\in\, 2\pi\, \mathbb{Z}\,,
\end{equation*}
which is the quantization condition (\ref{Magic}). In this case, we have the
additional requirement on the residues
\begin{equation}
-F_++F_-=\mathcal{A}_{\mathrm{on-shell}}\,.\,  \label{MagicReal}
\end{equation}

Consider this last constraint in the case of the off--shell eikonal
amplitude given in example (\ref{example}). The residue functions $f_{\pm }$
are then explicitly given by
\begin{equation*}
f_{\pm }\left( \sigma \right) =\mp a_{\pm }\frac{\sigma ^{2}}{4p}\,,
\end{equation*}
and we conclude that $F_{\pm }=\mp a_{\pm }\mathcal{A}_{\mathrm{on-shell}}$.
Therefore, condition (\ref{MagicReal}) implies that $a_{+}+a_{-}=1$, exactly
as in (\ref{exCondition}).

A few comments are in order. Firstly note that, in the regime of interest,
all kinematical invariants are much smaller than $M/J$. Assuming that the three
dimensional geometry comes from a ten dimensional string compactification
with Planck and string masses, respectively given by $M_{10}$ and $M_{s}$,
we see that string and Ka\l u\.{z}a--Klein effects are irrelevant when
$JM_{s}\gg M$ and when $J^{7/8}M_{10}\gg M$, which is always true for large charge $J$.

Secondly, suppose we use, instead of the full eikonal amplitude
$\mathcal{A}$, only the tree level result $-4s^{2}/tM$ for
gravitational scattering. Then, the single contribution to the
two--point function comes from the pole of the winding propagator,
violating the reality condition (\ref{singleReal}). This one--loop
violation of unitarity is analogous to the one found in the
computation of the one--point function in section \ref{OnePF},
where the tadpole $\Gamma \left( p\right) $ in (\ref{Four2}) has
support only for $p<0$.

Thirdly, within the eikonal approximation we cannot have a
violation of the reality of $\Gamma_{p_+,\,p_-}$ due to
intermediate on--shell lines. In fact, the eikonal interaction is
essentially transverse \cite{Levy:1969cr}, with no $k_+$ exchange.
Therefore, the scalar lines in Figure \ref{FigEikonal} have a
fixed value of the $K_+$ momentum. It is then impossible to cut
the graph so that all the cut propagators have positive $K_+$
momentum flowing from one part of the graph to the other.

Finally, note that, although the amplitude (\ref{example}) is the
simplest off--shell generalization of the on--shell eikonal
result, it is quite remarkable that we obtain the same constraint
on the coefficients $a_{\pm }$ by imposing the condition
(\ref{MagicReal}) on the residues. This fact clearly deserves a
more thorough investigation, which we leave for future work.

\section{Extending the results to off--shell external states\label{Extension}}

We now generalize the results of the last section to the case of
non--vanishing $\lambda ,\lambda ^{\prime }$. More precisely, we shall work
in the limit when $Ep_{+} \gtrsim \lambda,\lambda^{\prime}$ satisfying
\begin{equation*}
Ep_{+},\lambda ,\lambda ^{\prime }\ll \left( E/w\right) ^{2}\,,
\end{equation*}
where $w$ is the winding number of the loop propagator in Figure
\ref {FigOneLoop}. The first expression in equation
(\ref{saddleintegral}) is
still correct, but now the $\delta $--functions fix $q_{+}=2p/w$ and $%
q=(\lambda -\lambda ^{\prime })/\left( 2wp_{+}\right) $. Therefore we have
that
\begin{eqnarray*}
\Gamma _{w}\left( \mathbf{k,k}^{\prime }\right) =-\frac{T}{8}\;
e^{\,\frac{i}{E}\frac{w^{2}p}{12}} \int\, \frac{dq_{-}}{2\pi i}\;
\frac{1}{pwq_{-}-\frac{\left(\lambda -\lambda^{\prime}\right)^{2}}{16\,p_{+}^{\,2}}
+i\epsilon }\;\times \ \ \ \ \ \ \ \ &&\\
\times\;e^{\,\frac{i}{E}\,wq_{-}}\;
\mathcal{A}\left(\Omega^{-w/2}\mathbf{q,k,k}^{\prime},-\Omega^{w/2}\mathbf{q}\right)\,,&&
\end{eqnarray*}
with
\begin{eqnarray*}
\Omega ^{-w/2}\mathbf{q} &=&\left( \frac{2p}{w},q_{-}-\frac{\lambda -\lambda
^{\prime }}{4p_{+}}+\frac{wp}{4},\frac{\lambda -\lambda ^{\prime }}{2wp_{+}}-p\right) \,, \\
-\Omega ^{w/2}\mathbf{q} &=&-\left( \frac{2p}{w},q_{-}+\frac{\lambda
-\lambda ^{\prime }}{4p_{+}}+\frac{wp}{4},\frac{\lambda -\lambda ^{\prime }}{2wp_{+}}+p\right) \,.
\end{eqnarray*}
An approximation similar to the one used in equation (\ref{sut}) gives the
Mandelstam invariants
\begin{eqnarray*}
s &\simeq &\frac{4p}{w}\left( q_{-}+p_{-}\right) -\frac{\left( \lambda
-\lambda ^{\prime }\right) ^{2}}{4w^{2}p_{+}^{\,2}}\,, \\
t &=&-4p^{2}\,, \\
u &\simeq &\frac{4p}{w}\left( q_{-}-p_{-}\right) -\frac{\left( \lambda
-\lambda ^{\prime }\right) ^{2}}{4w^{2}p_{+}^{\,2}}\,.
\end{eqnarray*}
Just as for the on--shell case, the expression for
$\Gamma _{w}\left(\mathbf{k,k}^{\prime }\right)$ is invariant under $p\leftrightarrow -p$,
aside from conjugation of the phase. Moreover,
$\Gamma _{w}\left( \mathbf{k,k}^{\prime }\right)$ is invariant under $w\leftrightarrow -w$ and
$\lambda \leftrightarrow \lambda ^{\prime }$. Thus we arrive at the final
expression
\begin{equation}
\Gamma _{p_{+},p_{-}}\left( \lambda ,\lambda ^{\prime }\right)\, \simeq\,
e^{\,\frac{i}{E}\frac{\left( \lambda -\lambda ^{\prime }\right) ^{2}}{16p_{+}^{\,2}p}}\,
\sum_{w\neq 0}\,c_{w}\,\Gamma_{w}^{+}
\,+\,
e^{-\frac{i}{E}\frac{\left( \lambda-\lambda^{\prime}\right)^{2}}{16p_{+}^{\,2}p}}\,
\sum_{w\neq 0}\,c_{w}^{\,\star}\,\Gamma _{w}^{-}\;,
\label{offShellEx}
\end{equation}
where the constant $c_{w}$, given by (\ref{constFinal}), is unchanged from
the on--shell case and where
\begin{equation*}
\Gamma _{w}^{\pm }=\int \frac{ds}{2\pi i}\;
\frac{e^{\pm \frac{i}{4}\frac{w^{2}}{pE}s}}{\left|w\right|s-4pp_{-}+i\epsilon}\;
\mathcal{A}\left(s,t=-4p^{2},u=s-\frac{8pp_{-}}{\left| w\right| }\right)\,.
\end{equation*}
The expression for $\Gamma _{w}^{\pm }$ is formally identical to the
on--shell expression (\ref{AmpInt}), but yields a result depending on the
external off--shell masses, due to the implicit dependence of the momentum $p$
and of the amplitude $\mathcal{A}$ on $\lambda ,\lambda ^{\prime}$. As for
the on--shell case, the reality condition (\ref{realGamma}), together with
the fact that $\Gamma _{p_{+},p_{-}}\left( \lambda ,\lambda ^{\prime}\right)$
is symmetric in $\lambda ,\lambda ^{\prime }$, implies that
$\left(\sum c_{w}\,\Gamma_{w}^{+}\right)^{\star}=\sum c_{w}^{\,\star}\,\Gamma_{w}^{-}$.
Under the usual assumption regarding the separation of
different winding modes, we have that $\left( \Gamma _{w}^{+}\right) ^{\star
}=\,\Gamma _{w}^{-}\,$.

\subsection{Particle states}

To conclude the discussion of the two--point function, we wish to analyze
the single particle wave function $\Psi(\lambda)$, which solves the linearized
quantum equation of motion
\begin{equation}
\int d\lambda ^{\prime }\;
K\left(\lambda,\lambda^{\prime}\right)\;
\Psi\left( \lambda ^{\prime }\right) =0\,.
\label{singlePartState}
\end{equation}
The kernel $K$ is given by the full propagator (\ref{kernel})
\begin{equation*}
K\left( \lambda ,\lambda ^{\prime }\right) \,=\,
\left( \lambda +i\epsilon \right) \;
\delta \left( \lambda -\lambda ^{\prime }\right) +
\frac{1}{16\pi^{2}E}\;\Gamma_{p_{+},p_{-}}\left(\lambda,\lambda^{\prime}\right) \,.
\end{equation*}
In order to analyze (\ref{singlePartState}) we need to compute
$\Gamma _{p_{+},p_{-}}\left( \lambda ,\lambda ^{\prime }\right) $
for all values of $\lambda,\lambda^{\prime}$. However, we shall
use, for the discussion below, the results of the previous
section, which are strictly valid only for small values of
$\lambda $, $\lambda ^{\prime }$. Therefore the following analysis
should be considered heuristic and qualitative.

Equation (\ref{singlePartState}) is reminiscent of a scattering theory
problem and is formally solved by
\begin{equation}
\Psi =\phi -\frac{1}{16\pi ^{2}E}\;\frac{1}{\lambda \,\mathbf{1}+i\epsilon }\;
\Gamma _{p_{+},p_{-}}\Psi \,,  \label{LS}
\end{equation}
where $\phi(\lambda) =\delta(\lambda)$ is the
solution to the free equation of motion. Using (\ref{constFinal}) and (\ref
{onShellA}), we note that the amplitude $\Gamma _{p_{+},p_{-}}/E$ is of
order $\left( E/M\right) \left( E/p\right) ^{3/2}$ or $J^{-5/2}\left(
M/p\right) ^{3/2}$. Therefore, for large $J$ we can solve (\ref{LS}) in
powers of $\Gamma _{p_{+},p_{-}}$, with the leading ``Born'' term given by
\begin{equation*}
\Psi(\lambda) \simeq \delta(\lambda)-\frac{1}{16\pi^{2}E}\;
\frac{1}{\lambda \,+i\epsilon }\;\Gamma_{p_{+},p_{-}}\left(\lambda,0\right) \,.
\end{equation*}
In terms of $\Psi(\lambda)$, the dependence of the wave function
on the transverse coordinate $y$ is given by
\begin{equation*}
\Psi(y) =2\pi \left(\frac{2E}{\left|p_{+}^{\,}\right|}\right)^{\frac{1}{3}}\int d\lambda\;
\Psi(\lambda)\,\mathrm{Ai}(-z)\,.
\end{equation*}
The normalization has been chosen for later convenience and we recall that
\begin{equation}
z=\left( 2Ep_{+}^{\,2}\right) ^{\frac{1}{3}}\left( y+\frac{%
2p_{+}p_{-}-\lambda }{2Ep_{+}^{\,2}}\right) \,.  \label{zeq}
\end{equation}
Using the integral representation for the Airy function $\mathrm{Ai}$, we
can write $\Psi(y)$ explicitly as
\begin{equation*}
\Psi(y) =\frac{1}{\left| p_{+}\right| }\int d\lambda\, dk\;
\Psi(\lambda) \;
e^{\,\frac{i}{2Ep_{+}^{\,2}}
\left[\left(2Ep_{+}^{\,2}y+2p_{+}p_{-}-\lambda \right) k-\frac{k^{3}}{3}\right]}\,.
\end{equation*}

Let us now use the results of section \ref{Extension}. In particular, the
leading contribution to the $\lambda $--dependence of $\Gamma _{p_{+},p_{-}}$
comes from the implicit dependence of $p$ on $\lambda $ within the constant $%
c_{w}$ in equation (\ref{offShellEx}). Expanding the phase to linear order
in $\lambda $, and neglecting any other $\lambda $--dependence, the
expression for $\Gamma _{p_{+},p_{-}}(\lambda,0)$ is given by
\begin{equation*}
\frac{1}{16\pi ^{2}E}\;\Gamma _{p_{+},p_{-}}\left( \lambda ,0\right)\, \simeq\,
\frac{\alpha }{2\pi i}\;e^{-i\,\frac{p}{2Ep_{+}^{\,2}}\,\lambda }-
\frac{\alpha^{\star }}{2\pi i}\;e^{\,i\,\frac{p}{2Ep_{+}^{\,2}}\,\lambda }\,,
\end{equation*}
where we compute $p=\sqrt{2p_{+}p_{-}}$ and the constant $\alpha
=i\left( 8\pi E\right)^{-1}\sum_{w\neq 0}c_{w}\,\Gamma_{w}^{+}$ at
$\lambda=0$. The other term in the phase linear in $\lambda $ is
$w^{2}\lambda /\left( Ep\right) $ and is subleading since
$p_{+}\ll E/w^{2}$. Using the fact that $\int
d\lambda\,e^{-i\lambda\eta}\left(\lambda+i\epsilon\right)=-2\pi
i\,\theta(\eta)$, we can write the position space wave function as
\begin{equation*}
\Psi(y) \simeq \frac{1}{\left| p_{+}\right| }\int dk\,
\Big(1+\alpha\,\theta(k+p) -\alpha^{\star}\theta(k-p)\Big)\,
e^{\,\frac{i}{2Ep_{+}^{\,2}}\left[ \left( p^{2}+2Ep_{+}^{\,2}y\right)
k-\frac{k^{3}}{3}\right] }\,\,.
\end{equation*}
This integral can be computed using the saddle point approximation
for $\left( 2Ep_{+}^{\,2}\right) ^{-\frac{2}{3}}\left(
p^{2}+2Ep_{+}^{\,2}y\right) \gg 1$. In particular, this
approximation is valid around $y\sim 0$  since $p^{3}\gg
Ep_{+}^{\,2}\,$. For $y\sim 0$ the two saddle points at $k=\pm
\sqrt{p^{2}+2Ep_{+}^{\,2}y}$ are around $\pm p$ and the $\theta
$--functions create discontinuities. More precisely, we obtain
\begin{eqnarray*}
&&\left( 1+\alpha -\alpha ^{\star}\right) g_{+}(y) +g_{-}(y) \,,
\ \ \ \ \ \ \ \ \ \ \ \ \ \ \ \ \ \ \ \ \ \ (y>0)  \\
&&\left( 1+\alpha \right) \big[ g_{+}(y) +g_{-}(y)\big] \,,
\ \ \ \ \ \ \ \ \ \ \ \ \ \ \ \ \ \ \ \ \ \ \ \ \ \ \,(y<0)
\end{eqnarray*}
where $g_{\pm}(y)$ are the left and right moving waves
\begin{eqnarray*}
g_{\pm}(y)  &=&\sqrt{\mp 2\pi iE}\;
\left(p^{2}+2Ep_{+}^{\,2}y\right) ^{-1/4}\;
e^{\pm \frac{i}{3Ep_{+}^{\,2}}
\left(p^{2}+2Ep_{+}^{\,2}y\right) ^{3/2}}\;, \\
&\simeq &2\pi \left( \frac{2E}{\left| p_{+}\right| }\right)^{1/3}\;
\frac{\mathrm{Ai}\left( -z\right) \mp i\,\mathrm{Bi}\left( -z\right) }{2}\;,
\end{eqnarray*}
with $z$ given by (\ref{zeq}) for $\lambda =0$. The interaction term $%
\Gamma _{p_{+},p_{-}}$ creates a discontinuous behavior around
$y=0$, consistent with the conjectured dual description in terms
of an orientifold plane \cite{Cornalba:2002nv,Cornalba:2003kd}.
Due to the crudeness of the approximations involved, these ideas
clearly require a more thorough analysis.

\section{Conclusions}

The propagation of quantum fields in a geometry with closed causal curves is not
unitary order by order in the coupling constant. This breakdown of unitarity
arises from the interactions of external states with on--shell particle loops
that wind the closed causal curves. In this paper we showed that, for a orbifold
of three dimensional flat space with closed causal curves, unitarity can be restored for specific
values of the Newton coupling constant. The results relied on the eikonal approximation
to particle scattering, in a kinematic regime where graviton exchange is dominant.
Since in the eikonal approximation one is able to resum the perturbative series expansion,
this approximation provides a window to new effects in quantum gravity, such as the quantization
of the Newton constant from the unitarity requirement.
Alternatively, for fixed  coupling constant, one can see this condition as quantizing the orbifold geometry,
consistently with the quantization of the extremal BTZ black hole angular momentum. Given that
a similar condition holds for generic BTZ black holes, the results of this paper may
have profound implications for the dynamics of particle states inside black hole horizons.

We have used the off--shell extension of the eikonal amplitude for
$2\rightarrow 2$ scattering in flat space. Although the explicit
form of this amplitude is not known, we have only assumed that
such extension has the pole structure of the on--shell amplitude,
compatibly with the symmetries of the amplitude itself. Clearly,
it would be desirable to explicitly compute the off--shell
extension of the amplitude and to check if the additional
requirement (\ref{MagicReal}) is satisfied. In the usual flat space
computation, one needs to deal with IR divergences, which in
principle can be eliminated by considering eikonal scattering in
$\mathrm{AdS}_{3}$. We think this is an interesting direction of
research to pursue, since new connections between quantum gravity
in AdS spaces and dual CFT's may be derived.

Based on the particle effective action derived from the orbifold
two--point amplitude for off--shell external states, we gave in
the final section a heuristic argument showing that particle
states are changed precisely at the chronological singularity.
This analysis shows that, at finite values of the coupling,
particle states are very different from the free particle
wave--functions of section \ref{wavefunctions}. In fact, free
particle states are unstable beyond the chronological singularity.
Hence, for zero coupling constant, the condensation of these
fields  will change the geometry in the pathological region of
space--time \cite{Costa:2005ej}. It is plausible that, at the
end--point of this transition, the coupling constant is fixed at
some fixed value and the geometry is that of a wall placed at the
chronological singularity. This picture is suggested by a sequence
of string dualities that maps the orbifold geometry to an
orientifold 8--plane and the quantization condition (\ref{Magic})
to the quantization of RR charge \cite{Cornalba:2003kd}. Our
computation of the particle effective action, which sees finite
coupling constant effects, is also consistent with this picture.

\newpage\

\begin{center}
\textbf{Acknowledgments}
\end{center}

The authors wish to thank M. Bianchi, M. Ciafaloni, G. 't Hooft,
A.~Sagnotti, Y. Stanev and G. Veneziano for useful discussions and correspondence.

The authors thank the hospitality of the {\em Fields Institute for
Research in Mathematical Studies}, where this work was completed.
This work was supported in part by INFN, by the MIUR--COFIN
contract 2003--023852, by the EU contracts MRTN--CT--2004--503369,
MRTN--CT--2004--512194 and MERG--CT--2004--511309, by the INTAS
contract 03--51--6346, by the NATO grant PST.CLG.978785 and by the
FCT contracts POCTI/FNU/38004/ 2001 and  POCTI/FP/FNU/50161/2003.
L.C. is supported by the MIUR contract ``Rientro dei cervelli''
part VII. \emph{Centro de F\'{i}sica do Porto} is partially funded
by FCT through the POCTI programme.

\section*{Appendix A}

We focus on the extremal black hole, with $r_{+}=r_{-}$. Writing $\mathrm{AdS%
}_{3}$ as the surface
\begin{equation*}
-\left( z^{0}\right) ^{2}+\left( z^{1}\right) ^{2}+\left( z^{2}\right)
^{2}-\left( z^{3}\right) ^{2}=-\ell ^{2}
\end{equation*}
the BPS black hole is given by $\mathrm{AdS}_{3}/e^{\kappa }$, where
\begin{equation*}
\kappa =\frac{i}{\sqrt{2}}\left( J_{02}+J_{12}\right) +\frac{i}{\sqrt{2}}
\frac{\left( 2\pi r_{+}\right) ^{2}}{\ell ^{2}}\left(
J_{02}-J_{12}+J_{03}-J_{13}\right)
\end{equation*}
with $iJ_{\mu \nu }=z_{\mu }\partial _{\nu }-z_{\nu }\partial
_{\mu }$ the generators of the $SO\left( 2,2\right) $ isometry
group. Parameterizing $\mathrm{AdS}_{3}$ with coordinates $x^{a}$
($a=0,1,2$)
\begin{eqnarray*}
&&z^{0}+iz^{3} =-i\,e^{\,\frac{ix^{0}}{\ell }}\,\sqrt{\ell ^{2}+\left(
x^{1}\right) ^{2}+\left( x^{2}\right) ^{2}}\,, \\
&&z^{1}+iz^{2} =x^{1}+ix^{2}\,,
\end{eqnarray*}
we recover, in the $\ell \rightarrow \infty $ limit, flat Minkowski space
$\mathbb{M}^{3}$ with isometry group $ISO\left( 1,2\right) $. Moreover, the
generators $J_{\mu \nu }$ converge to the generators $K_{a},L_{ab}$ of
$ISO\left( 1,2\right)$, according to
\begin{equation*}
J_{ab}\rightarrow L_{ab}\,,\ \ \ \ \ \ \ \ \ \ \ \ \ \
\frac{1}{\ell }\,J_{3a}\rightarrow K_{a}\,.
\end{equation*}
In the limit $\ell \rightarrow \infty $, keeping the energy scale
\begin{equation*}
E=\frac{\ell }{\left( 2\pi r_{+}\right) ^{2}}
\end{equation*}
fixed, we arrive at the orbifold $\mathbb{M}^{3}/e^{\kappa }$ with
\begin{equation*}
\kappa =\frac{i}{\sqrt{2}}\left( L_{02}+L_{12}\right) +
\frac{i}{\sqrt{2}}\,E^{-1}\left( K_{0}-K_{1}\right)\,.
\end{equation*}
Introducing light--cone coordinates $x^{\pm }=\left( x^{0}\pm x^{1}\right) /%
\sqrt{2}$ and $x=x^{2}$, the metric on the covering space is
\begin{equation*}
ds^{2}=-2dx^{+}dx^{-}+dx^{2}\,,
\end{equation*}
and $\kappa $ simplifies to
\begin{equation*}
\kappa =i\left( L_{+x}+E^{-1}K_{-}\right)\,.
\end{equation*}

\section*{Appendix B}

We consider equation (\ref{MagicEQ}) for different values $w$, with $\sigma
=s-u=8\Lambda p_{-}/w^{2}$ and $2p=\sqrt{-t}=2\Lambda /\left| w\right| =%
\sqrt{\sigma \Lambda /2p_{-}}$ (we consider the case $p_{-}>0$ for
concreteness). Introduce, for notational convenience, the phase
\begin{equation*}
\zeta =e^{\,i\frac{M}{2E}}
\end{equation*}
and the constant $c=8\Lambda p$. Then (\ref{MagicEQ}) reads
\begin{equation}
-\zeta ^{w^{2}}F_+\left(c/w^{2}\right)+\zeta ^{-w^{2}}F_-\left(c/w^{2}\right)
=\mathcal{A}\left(c/w^{2}\right)\,,  \label{APP3}
\end{equation}
where we denote $\mathcal{A}\equiv \mathcal{A}_{\mathrm{on-shell}}$ for
brevity.

We wish to show, under minimal regularity assumptions, that (\ref{APP3}) can
be satisfied for all values of $w$ only if $\zeta =1$. To this end, we will
only need to assume that $F_{\pm},\mathcal{A}\in \mathcal{F}$, where $%
\mathcal{F}$ is a class of continuous real--valued functions of a real
variable defined on intervals $\left( 0,a\right) $, $a>0$, with the
following properties:

\begin{enumerate}
\item[$i)$]  if $f\in \mathcal{F}$, then either $f=0$ or $f(\sigma )\neq 0$
for $\sigma $ sufficiently close to $0$.

\item[$ii)$]  if $f,g\in \mathcal{F}$ and $\{\sigma _{w}\}$ is a sequence of
positive reals converging to zero such that $f(\sigma _{w})/g(\sigma _{w})$
is bounded, then $f(\sigma )/g(\sigma )$ converges for $\sigma \rightarrow 0$.

\item[$iii)$]  if $f,g\in \mathcal{F}$, then $f+g\in \mathcal{F}$.
\end{enumerate}

An example of a class of functions satisfying $i)$, $ii)$ and $iii)$ is the one of
finite sums of terms $x^{a}\log ^{n}(\sigma )\varphi (\sigma )$, where $a$
is a real number, $n$ is an integer and $\varphi $ is holomorphic on a
neighborhood of the origin in $\mathbb{C}$ and real on the real axis. This
example applies in particular to the case relevant for this paper in section
\ref{QCFU}, thanks to general properties of analyticity of scattering
amplitudes.

We first assume that $\zeta \neq \pm 1$. As shown in section \ref{QCFU}, the
function $\mathcal{A}$ is nowhere vanishing for $\sigma \neq 0$. Therefore,
if $F_+=0$, reality of the functions $F_-,\mathcal{A}$ and equation (\ref
{APP3}) imply that $F_-$ vanishes arbitrarily close to $0$, so that $%
F_-=0$ by $ii)$. A similar reasoning applies if $F_-=0$. Therefore we can
assume, by $i)$, that $F_{\pm },\mathcal{A}\neq 0$ in the range $0<\sigma
<1/W^{2}$ for some $W\in \mathbb{N}$. Since $\zeta \neq \pm 1$, there are
arbitrarily large values of $w$ such that $\zeta ^{w^{2}}$ is not real. We
denote with $S\in \mathbb{N}$ the set of these values of $w$ which are
greater than $W$. Equating imaginary parts in condition (\ref{APP3}) shows
that, for $w\in S$, $F_+\left(c/w^{2}\right)=-F_-\left(c/w^{2}\right)$,
and hence (\ref{APP3}) reduces to
\begin{equation}
2\func{Re}(\zeta ^{w^{2}})F_-\left(c/w^{2}\right)
=\mathcal{A}\left(c/w^{2}\right)\,. \ \ \ \ \ \ \ \ \ \ \ \ \ \
\left( w\in S\right)  \label{APP4}
\end{equation}
It follows from condition $ii)$ that the ratios
$\pm \mathcal{A}(\sigma)/F_{\mp }(\sigma )$ converge for $\sigma \rightarrow 0$ (to the same
limit). Now we distinguish two cases:
\begin{enumerate}
\item[a)]  $\zeta $ is a $k$--th root of unity. If $w\pm 1$ is a multiple of
$k$, then $\zeta ^{w^{2}}=\zeta $ and $w\in S$. Thus (\ref{APP4}) implies
that
\begin{equation*}
\pm \lim_{\sigma \rightarrow 0}\frac{\mathcal{A}(\sigma)}{F_{\mp}(\sigma)}=2\func{Re}(\zeta )\,.
\end{equation*}
On the other hand, condition (\ref{APP3}) for $w$ multiple of $k$ gives
\begin{equation*}
-\lim_{\sigma \rightarrow 0}\frac{F_-(\sigma) }{\mathcal{A}(\sigma )}+
\lim_{\sigma \rightarrow 0}\frac{F_+(\sigma) }{\mathcal{A}(\sigma )}=1
\end{equation*}
which is incompatible with the above, as $\zeta \neq 1$, unless $\zeta =\pm
i $. In this last case, however, (\ref{APP4}) shows that $\mathcal{A}\left(1/w^{2}\right)=0$
for all odd $w$, which contradicts the result of section \ref{QCFU}.

\item[b)]  $\zeta $ is not a root of unity. In this case $\zeta
^{w^{2}}$ is \textit{never} real for $w\neq 0$ and (\ref{APP4})
shows that the sequence $\left\{\func{Re}\left(\zeta
^{w^{2}}\right)\right\}$ converges to some limit $\cos \theta $.
Note, however, that $\zeta ^{\left( w+1\right) ^{2}}=\zeta
^{w^{2}}\zeta ^{2w+1}$ and $\zeta ^{2w+1}$ is dense on the unit
circle as we vary $w$. Therefore we can choose $w$ large enough so
that $\func{Re}\zeta ^{w^{2}}$ is arbitrarily close to $\cos
\theta $, and so that $\zeta ^{2w+1}$ is arbitrarily close to any
chosen number $\xi $ on the unit circle. By choosing $\xi $
different from $e^{\pm 2i\theta }$, we can then make
$\func{Re}\zeta ^{\left( w+1\right) ^{2}}$ differ from $\cos
\theta $ by a finite amount for arbitrarily large values of $w$,
therefore showing that the sequence $\left\{\func{Re}\left(\zeta
^{w^{2}}\right)\right\}$ does not converge.
\end{enumerate}

We finally deal with the case $\zeta =-1$. From (\ref{APP3}) we deduce that
\begin{equation*}
-F_+\left(c/w^{2}\right)+F_-\left(c/w^{2}\right)=(-) ^{w}\mathcal{A}\left(c/w^{2}\right)\,.
\end{equation*}
Since $\mathcal{A}$ has constant sign near $0$, this relation implies that
$-F_+(\sigma )+F_-(\sigma)$ vanishes arbitrarily close to $0$ and is
non--vanishing. This is a contradiction by $iii)$.

\end{document}